\pdfoutput=1
\documentclass[12pt]{article}
\PassOptionsToPackage{unicode}{hyperref}
\PassOptionsToPackage{hyphens}{url}
\PassOptionsToPackage{dvipsnames,svgnames,x11names}{xcolor}

\usepackage{amsmath,amssymb}
\usepackage{iftex}
\ifPDFTeX
  \usepackage[T1]{fontenc}
  \usepackage[utf8]{inputenc}
  \usepackage{textcomp} 
\else 
  \usepackage{unicode-math}
  \defaultfontfeatures{Scale=MatchLowercase}
  \defaultfontfeatures[\rmfamily]{Ligatures=TeX,Scale=1}
\fi
\usepackage{lmodern}
\ifPDFTeX\else  
\fi
\IfFileExists{upquote.sty}{\usepackage{upquote}}{}
\IfFileExists{microtype.sty}{
  \usepackage[]{microtype}
  \UseMicrotypeSet[protrusion]{basicmath} 
}{}
\makeatletter
\@ifundefined{KOMAClassName}{
  \IfFileExists{parskip.sty}{%
    \usepackage{parskip}
  }{
    \setlength{\parindent}{0pt}
    \setlength{\parskip}{6pt plus 2pt minus 1pt}}
}{
  \KOMAoptions{parskip=half}}
\makeatother
\usepackage{xcolor}
\makeatletter
\ifx\paragraph\undefined\else
  \let\oldparagraph\paragraph
  \renewcommand{\paragraph}{
    \@ifstar
      \xxxParagraphStar
      \xxxParagraphNoStar
  }
  \newcommand{\xxxParagraphStar}[1]{\oldparagraph*{#1}\mbox{}}
  \newcommand{\xxxParagraphNoStar}[1]{\oldparagraph{#1}\mbox{}}
\fi
\ifx\subparagraph\undefined\else
  \let\oldsubparagraph\subparagraph
  \renewcommand{\subparagraph}{
    \@ifstar
      \xxxSubParagraphStar
      \xxxSubParagraphNoStar
  }
  \newcommand{\xxxSubParagraphStar}[1]{\oldsubparagraph*{#1}\mbox{}}
  \newcommand{\xxxSubParagraphNoStar}[1]{\oldsubparagraph{#1}\mbox{}}
\fi
\makeatother

\usepackage{longtable,booktabs,array,makecell}
\usepackage{calc} 
\usepackage{etoolbox}
\makeatletter
\patchcmd\longtable{\par}{\if@noskipsec\mbox{}\fi\par}{}{}
\makeatother
\IfFileExists{footnotehyper.sty}{\usepackage{footnotehyper}}{\usepackage{footnote}}
\makesavenoteenv{longtable}
\usepackage{graphicx}
\makeatletter
\def\maxwidth{\ifdim\Gin@nat@width>\linewidth\linewidth\else\Gin@nat@width\fi}
\def\maxheight{\ifdim\Gin@nat@height>\textheight\textheight\else\Gin@nat@height\fi}
\makeatother
\setkeys{Gin}{width=\maxwidth,height=\maxheight,keepaspectratio}
\makeatletter
\def\fps@figure{htbp}
\makeatother

\makeatletter
\@ifpackageloaded{caption}{}{\usepackage{caption}}
\AtBeginDocument{%
\ifdefined\contentsname
  \renewcommand*\contentsname{Table of contents}
\else
  \newcommand\contentsname{Table of contents}
\fi
\ifdefined\listfigurename
  \renewcommand*\listfigurename{List of Figures}
\else
  \newcommand\listfigurename{List of Figures}
\fi
\ifdefined\listtablename
  \renewcommand*\listtablename{List of Tables}
\else
  \newcommand\listtablename{List of Tables}
\fi
\ifdefined\figurename
  \renewcommand*\figurename{Figure}
\else
  \newcommand\figurename{Figure}
\fi
\ifdefined\tablename
  \renewcommand*\tablename{Table}
\else
  \newcommand\tablename{Table}
\fi
}
\@ifpackageloaded{float}{}{\usepackage{float}}
\floatstyle{ruled}
\@ifundefined{c@chapter}{\newfloat{codelisting}{h}{lop}}{\newfloat{codelisting}{h}{lop}[chapter]}
\floatname{codelisting}{Listing}

\makeatother
\makeatletter
\@ifpackageloaded{caption}{}{\usepackage{caption}}
\@ifpackageloaded{subcaption}{}{\usepackage{subcaption}}
\makeatother

\ifLuaTeX
  \usepackage{selnolig}  
\fi
\usepackage[square]{natbib}
\usepackage{bookmark}

\IfFileExists{xurl.sty}{\usepackage{xurl}}{} 
\usepackage{xcolor}
\definecolor{mutedblue}{RGB}{30, 100, 200} 

\hypersetup{
  pdftitle={Title},
  pdfauthor={Author 1; Author 2},
  pdfkeywords={3 to 6 keywords, that do not appear in the title},
  colorlinks=true,
  linkcolor={mutedblue},
  filecolor={mutedblue},
  citecolor={mutedblue},
  urlcolor={mutedblue},
  pdfcreator={LaTeX via pandoc}}

\usepackage{setspace} 
\def\spacingset#1{\renewcommand{\baselinestretch}%
{#1}\small\normalsize} \spacingset{1}

\usepackage{nicefrac}
\usepackage{relsize}

\usepackage{hyperref}

\usepackage[capitalize,nameinlink]{cleveref}
\crefrangelabelformat{equation}{(#3#1#4--#5#2#6)}
\crefname{section}{section}{sections} 
\crefname{appendix}{appendix}{appendices}
\crefalias{appsec}{section}

\usepackage{chngcntr}

\def\spacingset#1{\renewcommand{\baselinestretch}%
{#1}\small\normalsize} \spacingset{1}

\usepackage{centernot}
\usepackage{amsthm}         
\usepackage{nicefrac}       
\usepackage{mathtools}      
\usepackage{amsbsy}         
\usepackage{amstext}        
\usepackage{thmtools}       
\usepackage{thm-restate}    

\begingroup
    \makeatletter
    \@for\theoremstyle:=definition,remark,plain\do{%
        \expandafter\g@addto@macro\csname th@\theoremstyle\endcsname{%
            \addtolength\thm@preskip\parskip
            }%
        }
\endgroup

\crefname{lemma}{Lemma}{Lemmas}
\Crefname{lemma}{Lemma}{Lemmas}
\crefname{theorem}{Theorem}{Theorems}
\Crefname{theorem}{Theorem}{Theorems}
\crefname{proposition}{Proposition}{Propositions}
\Crefname{proposition}{Proposition}{Propositions}
\crefname{cor}{Corollary}{Corollaries}
\Crefname{cor}{Corollary}{Corollaries}
\crefname{defn}{Definition}{Definitions}
\Crefname{defn}{Definition}{Definitions}
\crefname{assumption}{Assumption}{Assumptions}
\Crefname{assumption}{Assumption}{Assumptions}

\usepackage{arydshln}
\makeatletter
\def\adl@drawiv#1#2#3{%
        \hskip.5\tabcolsep
        \xleaders#3{#2.5\@tempdimb #1{1}#2.5\@tempdimb}%
                #2\z@ plus1fil minus1fil\relax
        \hskip.5\tabcolsep}
\newcommand{\cdashlinelr}[1]{%
  \noalign{\vskip\aboverulesep
           \global\let\@dashdrawstore\adl@draw
           \global\let\adl@draw\adl@drawiv}
  \cdashline{#1}
  \noalign{\global\let\adl@draw\@dashdrawstore
           \vskip\belowrulesep}}
\makeatother

\renewcommand{\epsilon}{\varepsilon}

\declaretheorem[style=plain,numberwithin=section,name=Theorem]{theorem}

\declaretheorem[style=plain,numberwithin=section,name=Proposition]{proposition}
\declaretheorem[style=plain,numberwithin=section,name=Corollary]{cor}

\declaretheorem[style=plain,numberwithin=section,name=Definition]{defn}

\declaretheorem[style=plain,numberwithin=section,name=Example]{example}
\declaretheorem[style=plain,numberwithin=section,name=Remark]{remark}

\newenvironment{example*}
 {\pushQED{\qed}\example}
 {\popQED\endexample}

\newcommand{\defeq}{\overset{\mathrm{def}}{=}}

\newcommand{\indsim}{\overset{\mathrm{ind}}{\sim}}

\newcommand{\exclude}{\backslash}

\DeclareMathOperator*{\logit}{logit}

\newcommand{\ms}{\mathsmaller}
\newcommand{\squish}[1]{\!#1\!}
\newcommand{\tm}{\squish{-}}

\newcommand{\tp}{\squish{+}}
\newcommand{\teq}{\squish{=}}

\newcommand{\ex}{\operatorname{\mathbb{E}}}

\usepackage{dsfont}

\newcommand{\TNB}{\mathrm{TNBbeta}}

\newcommand{\bodylabel}{\label}

\usepackage{natbib}
\usepackage[affil-it]{authblk}

\newcommand{\paperTitle}{\LARGE \bf The Triply-Randomized Negative Binomial Beta for Robust Regression and Conjugate Models of Bounded Support Data}
\author[1]{Jimmy Lederman}
\author[1,2]{Aaron Schein}

\affil[1]{Department of Statistics, University of Chicago}
\affil[2]{Data Science Institute, University of Chicago}
\date{}

\begin{document}

\title{\paperTitle}
\maketitle

\bigskip

\begin{abstract}
    
The beta distribution is the default choice of likelihood in many regression problems with a $[0,1]$-bounded support response despite its sensitivity to outliers, inability to accommodate exact zero observations, and a lack of closed-form conjugate priors. We address these shortcomings by introducing the \textit{triply-randomized negative binomial beta distribution}, denoted $\TNB(p,\,q,\,\varepsilon)$, parameterized by a median $p$, concentration parameter $q$, and boundary parameter $\varepsilon$ which permits positive density at $0$ and $1$. The TNBbeta arises by randomizing the parameters of a standard beta distribution with three dependent negative binomial random variables, each of whose complete conditional distribution we show is itself negative binomial. Moreover, connecting $p$ and $q$ to Gaussian latent variables with logit link functions yields closed-form updates via Pólya-gamma augmentation. Together, these properties yield simple auxiliary-variable Gibbs samplers for regression models of bounded-support data, which often outperform standard beta regression approaches in terms of effective sample size per second and held-out prediction, especially in the presence of outliers. In a case study of forest canopy cover, we demonstrate that this framework can easily incorporate spatial structure and exact zero observations. Overall, this work substantially expands the class of Bayesian models for $[0,1]$-bounded support data that can be fit efficiently.

\end{abstract}

\noindent%
\vfill

\newpage
\spacingset{1.8} 

\section{Introduction}

\begin{figure}
\begin{center}
\includegraphics[width=\linewidth]{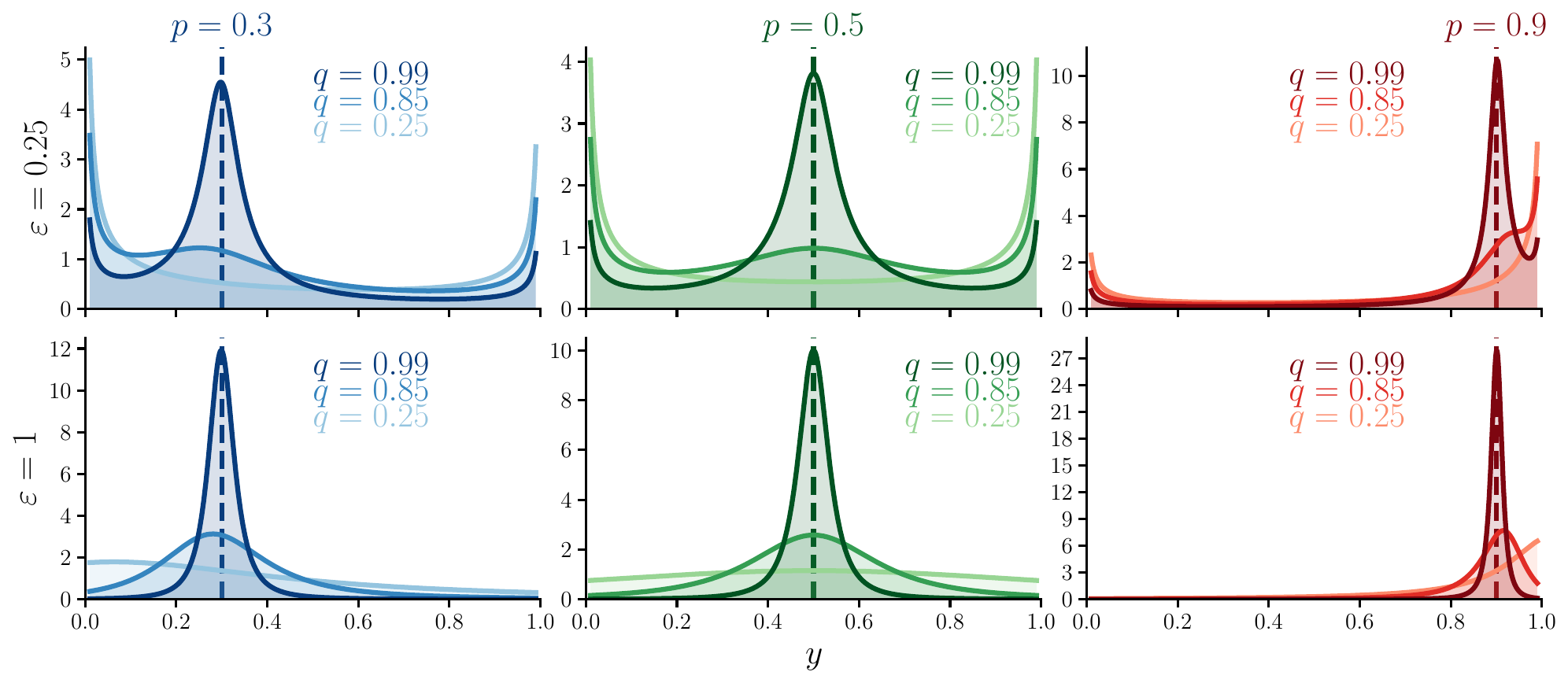}
\end{center}
\caption{\label{figure:pdf1}The TNBbeta family is highly flexible, with $p$ shifting the median, $q$ controlling the concentration around $p$, and $\varepsilon$ modulating boundary behavior.}
\end{figure}

This paper introduces a new family of probability distributions
on the unit interval which we call \textit{triply-randomized negative binomial beta} (\textit{TNBbeta}) \textit{distributions} with density function
\begin{equation}
\TNB(y;\,p,\,q,\,\varepsilon)
=
\textrm{beta}(y;\,\varepsilon,\varepsilon)
\left[
\tfrac{1-q}{y(1-y)}
\gamma(y,p)
\right]^{\varepsilon}
\big[1-4q\,\gamma(y,p)\big]^{-(\varepsilon + \nicefrac{1}{2})}
 \end{equation}

where
$\gamma(y,p)\defeq \tfrac{1}{4}\textrm{sech}^2\big(\nicefrac{\textrm{logit}(y)-\textrm{logit}(p)}{2}\big)$
is a similarity function between the log-odds of $y$ and $p$, bounded $0 \!<\! \gamma(y,p)\!\le\! 1$ and maximized at $y\teq p$. It is most natural to parameterize the TNBbeta by its \textit{median} $p$, which endows models  with interpretability and robustness to outliers. The parameter $q$ then controls the distribution's \textit{concentration} around $p$, while $\varepsilon$ governs its behavior at the boundaries, permitting positive density at 0 and 1 when $\varepsilon \teq 1$ and tri-modal density when $\varepsilon \!<\! 1$ (see~\Cref{figure:pdf1}). As we show, these properties make the TNBbeta highly appealing as a likelihood in models for $ [0,1]$-bounded support data.

The TNBbeta arises as the marginal of the following randomization of the standard beta
\begin{equation}\label{eq1}
    \begin{aligned}
     Y &\sim \textrm{beta}\big(\varepsilon \!+\! A \!+\! C,\,\, \varepsilon \!+\! B \!+\! C\big),\\
    A  \sim \textrm{NB}\big(\varepsilon \!+\! C,\, 1-&p\big),\quad B \sim \textrm{NB}\big(\varepsilon \!+\! C, \,p\big), \quad C \sim \textrm{NB}\big(\varepsilon,\,1\tm q\big).
    \end{aligned}
\end{equation}
This construction follows in a tradition of generalizing the beta distribution via discrete mixing~\citep{johnson_continuous_1995}. We establish a remarkable \emph{reverse conjugacy} property:
the complete conditional of each auxiliary negative binomial above is
itself negative binomial. 

These results enable efficient inference in TNBbeta regression models of the following form:
\begin{equation}\label{eq:pgmodel1}
    Y_i \sim \TNB(p_i,\,q_i,\,\varepsilon),\qquad p_i = \textrm{logit}^{-1}\big(X_i^\top \beta_p\big), \qquad q_i = \textrm{logit}^{-1}\big(X_i^\top \beta_q\big),
\end{equation}
where $\beta_p$ and $\beta_q$ are coefficients for the median and concentration. Concretely, such results enable conditionally conjugate posterior inference via a new Pólya-gamma augmentation scheme~\citep{polson_bayesian_2013}, thus also enabling structured Gaussian priors---e.g., for hierarchical, temporal, or spatial latent variables---in models for $[0,1]$-bounded data.

The standard beta distribution has long been the default choice of prior distribution for bounded support parameters in Bayesian analysis. Due to its ubiquity as a prior, the beta is also widely used as a likelihood in probabilistic models of $(0,1)$-bounded support data. Beta regression~\citep{ferrari_beta_2004}, for example, is widely used in ecology~\citep{damgaard_using_2019,douma_analysing_2019}, medicine~\citep{liu_review_2018}, and the social sciences~\citep{verkuilen_mixed_2012}. However, despite its convenience as a prior, the beta is challenging to work with as a likelihood due to its parameters entering special functions and its lack of any closed-form conjugate prior. As a result, parameter inference and estimation in these settings can become computationally prohibitive outside simple model classes, hampering the ability to encode complex prior structure.

Existing alternatives to the beta likelihood for $(0,1)$-bounded support data each entail substantial tradeoffs. Perhaps the most common approach is to ignore boundedness and utilize a likelihood with unbounded support, most commonly the Gaussian. This approach is prone to leakage of probability density outside the unit interval and often fails to capture asymmetries in bounded support data induced by the geometry of the unit interval~\citep{kieschnick_regression_2003}. Another option is to introduce a transformation that maps data to the real line and then model the transformed data
with an unbounded likelihood. However, parameter estimates in such models are not easily interpreted on the scale of the original response~\citep{cribari-neto_beta_2010}. The last approach is to build models with other distributions with unit interval support~\citep{gupta_handbook_2004,afuecheta_review_2025}, but no such distribution has seen widespread use as a likelihood due to a general lack of closed forms---such as probability densities, link functions, and conjugate priors---which obscure parameter interpretation and complicate tractable posterior inference.

We address these shortcomings with a closed-form,
median-parameterized likelihood whose conditionally conjugate properties we show lead to tractable parameter inference for a broad class of regression and latent variable models. Our main contributions are
as follows.
\begin{itemize}
\item We introduce the TNBbeta in~\Cref{sec:dist}
and demonstrate that its parameters are directly interpretable in regression models.
\item In~\Cref{sec:mixing}, we show that the TNBbeta arises from randomizing the standard beta with three negative binomial auxiliary 
variables, each admitting a form of conjugacy.

\item We use these conjugacy results to derive auxiliary-variable Gibbs samplers for TNBbeta regression models in~\Cref{sec:mcmc} . To do so, we introduce a new  P\'{o}lya-gamma augmentation scheme~\citep{polson_bayesian_2013} for $(0,1)$-bounded support distributions in~\Cref{sec:pglnb}.

\item We show that TNBbeta regression models are more robust to noise (\Cref{sec:robust}) and often fit better and more efficiently (\Cref{sec:bench}) than alternatives. In~\Cref{sec:trees}, we apply these models to 
ecological data with spatial structure and boundary observations.
\end{itemize}

\section{The Beta Distribution and its Generalizations}\label{sec:background}
Our approach builds on known properties and generalizations of the beta distribution.
\begin{defn}
    A beta random variable $Y \sim \mathrm{beta}(\alpha_1,\alpha_2)$ has probability density function 
    \begin{equation}\label{eq:beta}\mathrm{beta}(y;\,\alpha_1,\,\alpha_2) = \frac{y^{\alpha_1 - 1}(1 \tm y)^{\alpha_2 - 1}}{B(\alpha_1,\,\alpha_2)},
\end{equation}
with shape parameters $\alpha_1 > 0$, $\alpha_2 > 0$ and where $B(a,b) = \nicefrac{\Gamma(a)\Gamma(b)}{\Gamma(a+b)}$ is the beta function. An equivalent parameterization with mean $\mu \in  (0,1)$ and concentration parameter $\phi > 0$ is\begin{equation}\label{eq:parambeta}
    \mathrm{beta}\big(\textsc{mean}\teq \mu,\,\,\textsc{concentration}\teq\phi\big)\Longleftrightarrow\mathrm{beta}\big(\phi\mu,\,\,\phi (1\tm \mu)\big),
\end{equation}
\end{defn}

The beta function makes working with beta shape parameters cumbersome, prohibiting, for example, analytic marginalization. As~\cite{johnson_continuous_1995} note,
 \begin{quote}
 ``\textit{Compound beta} distributions may be formed by ascribing distributions to [$\alpha_1$ and $\alpha_2$]. However, such distributions have not been used much in applied statistical work \dots and continuous distributions [for $\alpha_1$ and $\alpha_2$] usually present analytical difficulties, owing to the presence of the beta function.''
 \end{quote}
 
 The difficulty of working with continuous distributions for the shape parameters motivates the use of discrete distributions, and~\cite{johnson_continuous_1995} list several choices of discrete mixing distribution which induce appealing analytic properties.
 
Consider a general discrete mixing distribution $\mathcal{DM}$ for auxiliary count $A$. The construction
\begin{equation}\label{eq:singlecount}
    A\sim \mathcal{DM}, \qquad  Y \mid A \teq a \sim \textrm{beta}\big(\alpha_1 \tp a,\,\alpha_2\big),
\end{equation}
often leads to an analytic marginal distribution for $Y$. In fact, many beta generalizations can be represented as compound beta distributions of this form for different choices of mixing distribution $\mathcal{DM}$. For example,~\cite{roy_binomial_1993} introduced the binomial mixture of beta distributions in which $\mathcal{DM}$ is binomial. Choosing $\mathcal{DM}$ to be Poisson yields the non-central beta distributions of the first and second kind~\citep{hodges_noncentral_1955,seber_non-central_1963}. 

One of the most prominent generalizations of the beta is the \textit{Libby-Novick beta distribution}~\citep{libby_multivariate_1982,chen_bayesian_1984}, which we define below. 

\begin{defn} \label{def:lnb}
    A Libby-Novick beta random variable $Y\sim\mathrm{LNbeta}(\alpha_1,\,\alpha_2,\,c)$ has density
    \begin{equation}
    \label{eq:lnb}
    \mathrm{LNbeta}(y;\,\alpha_1,
    \,\alpha_2,
    \,c) = \mathrm{beta}(y;\,\alpha_1,\,\alpha_2)\cdot\frac{c^{\alpha_1}}{\big(1-(1-c)y\big)^{\alpha_1+\alpha_2}},
\end{equation}
with shape parameters $\alpha_1 > 0$, $\alpha_2 > 0$ and tilting parameter $c > 0$. 
\end{defn}

Unlike most other beta generalizations, the LNbeta does not introduce additional special functions into its normalizing constant. A characterizing property of this distribution is its construction as the ratio of two gamma-distributed random variables---i.e., if $G_i \indsim \Gamma(\alpha_i,\beta_i)$ then $\nicefrac{G_1}{G_1+G_2} \sim \text{LNbeta}\big(\alpha_1,\,\alpha_2,\,\tfrac{\beta_1}{\beta_2}\big)$. Setting $\beta_1\teq \beta_2$  recovers the well-known representation of the standard beta as the ratio of two gamma random variables with equal rates.

Recent work~\citep{chabot_sur_2016,jones_noncentral_2021} has shown that the LNbeta also has an auxiliary-count representation as in~\Cref{eq:singlecount} in which $\mathcal{DM}$ is negative binomial
\begin{equation}\label{eq:nblnb1}
A \sim \mathrm{NB}(\alpha_1,\,q),\qquad Y \mid A\teq a \sim \mathrm{beta}\big(\alpha_1 \tp a,\,\alpha_2\big).
\end{equation}
Note that adding $A$ instead to the second shape parameter in~\Cref{eq:nblnb1} also yields a LNbeta marginal, but only if the negative binomial shape matches the second beta shape---i.e., $A \sim \textrm{NB}\big(\alpha_2,\,q\big)$.
The addition of $A$ to the first or second shape perturbs the marginal distribution towards $0$ or $1$, respectively.~\cite{jones_slew_2023} show that the inverse distribution for the auxiliary count $A$ in~\Cref{eq:nblnb1} is also negative binomial
\begin{equation}\label{eq:nblnb2}
    A \mid Y\teq y \sim \text{NB}\big(\alpha_1\tp\alpha_2,\,1\tm(1\tm q)y\big),
\end{equation}
demonstrating a form of conjugacy between a beta likelihood and negative binomial prior.

The Libby–Novick beta belongs to the broader family of what~\cite{jones_slew_2023} term the \textit{beta-generated continuous generalized hypergeometric distributions}. This family includes several notable beta generalizations, such as the confluent hypergeometric beta~\citep{gordy_computationally_1998} and the Gauss hypergeometric beta~\citep{armero_prior_1994}. Each distribution in this class admits an auxiliary-variable representation of the form~\Cref{eq:singlecount}, where $\mathcal{DM}$ is a discrete generalized hypergeometric distribution~\citep{kemp_wide_1968}. Moreover, the inverse distribution for $A$ in each case is also 
discrete generalized hypergeometric. \Cref{eq:nblnb1,eq:nblnb2} emerge as a simple instance of this general construction, corresponding to when the discrete generalized hypergeometric distribution reduces to negative binomial.

The representation in~\Cref{eq:singlecount} can be thought of as a single auxiliary-count construction because only one shape parameter is randomized. Though less common in the literature, some distributions on $(0,1)$ can be expressed via a \textit{bivariate} auxiliary-count representation,
\begin{equation}\label{eq:doublecount}
    \begin{aligned}
    (A,B) &\sim \mathcal{DM},
        \\Y \mid A \teq a,\, B \teq b &\sim \textrm{beta}\big(\alpha_1 \tp a,\,\alpha_2 \tp b\big),
    \end{aligned}
\end{equation}
where $\mathcal{DM}$ is now bivariate. In simple cases, $(A,B)$ are independent---i.e. $\mathcal{DM}(a,b) \equiv \mathcal{DM}_1(a)\cdot\mathcal{DM}_2(b)$. In other cases, $(A,B)$ are dependent under $\mathcal{DM}$, as in the case of the \textit{conditional doubly non-central beta}~\citep{ongaro_results_2015}, which arises when $A$ and $B$ are Poisson random variables which condition on a common value. When $A$ and $B$ are independent Poisson random variables, the marginal distribution of \Cref{eq:doublecount} is the \textit{doubly non-central beta distribution}~\citep{ongaro_results_2015}, with probability density
\begin{equation}
\label{eq:dncbeta}
\textrm{DNCbeta}(y;\alpha_1,\!\alpha_2,\!\lambda_1,\!\lambda_2) \teq
\textrm{beta}(y;\alpha_1,\!\alpha_2)\, e^{-\scriptstyle\frac{1}{2}(\lambda_1 \tp \lambda_2)}\Psi_2\big(\alpha_1\tp\alpha_2;\alpha_1,\alpha_2,\tfrac{1}{2}\lambda_1 y,\tfrac{1}{2}\lambda_2(1\tm y)\big)
\end{equation}
where $(\alpha)_{j}$ is the Pochhammer symbol and $\Psi_2$ is Humbert's confluent hypergeometric
\begin{equation}\label{eq:humbert}
\Psi_2 \big[\alpha; \gamma, \gamma'; x, y \big] = 
\sum_{j=0}^{\infty} \sum_{k=0}^{\infty} 
\frac{(\alpha)_{j+k} x^j y^k}{(\gamma)_j (\gamma')_k\, j!\, k!},
\quad x, y \ge 0.
\end{equation}
Though computable, this special function complicates inference for the ``non-centrality'' parameters $\lambda_1$ and $\lambda_2$---a difficulty shared 
by most distributions discussed in this section.

We next introduce a beta generalization whose 
auxiliary-count construction avoids these difficulties, yielding 
closed-form conjugacy and tractable, interpretable parameter inference.

\section{The Triply-Randomized Negative Binomial Beta}\label{sec:dist}

We introduce the $\TNB$, a novel beta generalization with interpretable parameters. We arrive at this distribution via a new negative binomial mixing distribution for the standard beta in~\Cref{sec:mixing}. We first study key $\TNB$ properties, deferring proofs to~\Cref{app:proof1}.

\newcommand{\propOurbetaNBBody}{%
Let $p\!\in\!(0,1)$, $q\!\in\!(0,1)$, and $\varepsilon\!>\!0$. A \textit{triply-randomized negative binomial beta} random variable $Y \sim \TNB\big(p,\,q,\,\varepsilon\big)$ has probability density
\begin{equation}\bodylabel{eq:pdf}
\TNB(y;\,p,\,q,\,\varepsilon)
=
\mathrm{beta}(y;\,\varepsilon,\varepsilon)
\left[
\tfrac{1-q}{y(1-y)}
\gamma(y,p)
\right]^{\varepsilon}
\big[1-4q\,\gamma(y,p)\big]^{-(\varepsilon + \nicefrac{1}{2})}
\end{equation}
where $\gamma(y,p)$ is a similarity function of the log-odds of $y$ and $p$ as follows
\begin{equation}\bodylabel{eq:g}\gamma(y,p)\defeq \tfrac{1}{4}\mathrm{sech}^2\big(\nicefrac{\mathrm{logit}(y)-\mathrm{logit}(p)}{2}\big) = \frac{y(1-y)p(1-p)}{(p(1-y) + (1-p)y)^{2}}
\end{equation}
}
\begin{defn}[\textbf{The TNBbeta}]\label{prop:ourbetaNB} \propOurbetaNBBody
\end{defn}
\begin{figure}
\begin{center}
\includegraphics[width=\linewidth]{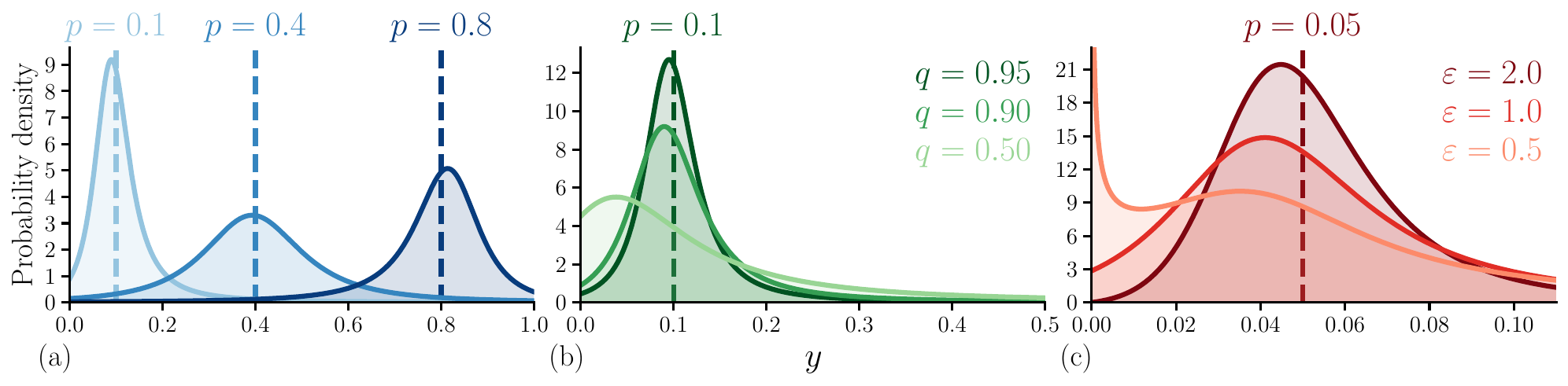}
\end{center}
\caption{\label{fig:pdfs}(a) The median $p$ shifts the centrality of $\TNB\big(y;\;p,\,0.9,\,1\big)$. (b) As $q$ increases, $\TNB\big(y;\;0.1,\,q,\,1\big)$ concentrates around its median $p\teq.1$. (c) The boundary density $\TNB\big(0;\;0.05,\,.9,\,\varepsilon\big)$  can diverge to $\infty$, be positive finite, or approach $0$ depending on $\varepsilon$.}
\end{figure}

Unlike most beta generalizations, the TNBbeta has closed-form density, and its parameters $p$ and $q$ do not enter the beta function.~\Cref{fig:pdfs}a shows that $p$ controls the TNBbeta centrality, and~\Cref{fig:pdfs}b shows that $q$ dictates the concentration around $p$, as stated below. 

\newcommand{\thmMedianTwoTitle}{\textbf{TNBbeta median-concentration}}
\newcommand{\propMedianTwoBody}{%
Let $Y \sim \mathrm{TNBbeta}\big(p,\,q,\,\varepsilon\big)$. Then $p$ and $q$ are Fisher orthogonal and separately control the median and concentration such that
\begin{enumerate}
\item $\mathrm{median}(Y) = p$
\item $\lim_{q\to 1}  \mathrm{\TNB}(p;\,p,\,q,\,\varepsilon) = \infty$
\item The Fisher information matrix satisfies $\mathcal{I}_{pq}(p,\,q,\,\varepsilon) = \ex\!\left[-\frac{\partial^2 \log \TNB(Y;\,p,\,q,\,\varepsilon)}{\partial p\,\partial q}\right] = 0$.
\end{enumerate}
}
\begin{theorem}[\thmMedianTwoTitle]
\label{theorem:tnbebta} \propMedianTwoBody
\end{theorem}
\Cref{theorem:tnbebta} mirrors the mean-concentration parameterization of the standard beta in \Cref{eq:parambeta}. However, the beta mean and concentration are not Fisher orthogonal, so their maximum likelihood estimates are dependent~\citep{ferrari_beta_2004}. Orthogonality ensures that inference on the median does not interfere with inference on the concentration.

The remaining parameter $\varepsilon$ governs the modal and boundary behavior.~\Cref{fig:pdfs}c shows that when $\varepsilon \teq .5$ 
the TNBbeta can attain both interior and boundary modes. In contrast, when $\varepsilon \teq1$ or $\varepsilon\teq2$, there are no boundary modes and the density at the boundary is either finite and positive or approaches zero, respectively.~\Cref{prop:phi} formalizes each regime.

\newcommand{\propPhiTitle}{\textbf{Boundary behavior}}
\newcommand{\propPhiBody}{%
The TNBbeta boundary density depends on $\varepsilon$:
\begin{equation*}
\big(\TNB(0;\,p,\,q,\,\varepsilon),\;\TNB(1;\,p,\,q,\,\varepsilon)\big) = \begin{cases} (0,\;0) & \varepsilon > 1 \\ \big(\tfrac{(1-p)(1-q)}{p},\;\tfrac{p(1-q)}{1-p}\big) & \varepsilon = 1 \\ (\infty,\;\infty) & \varepsilon < 1 \end{cases}
\end{equation*}
}
\begin{proposition}[\propPhiTitle]
\label{prop:phi} \propPhiBody
\end{proposition}

\begin{remark}
  The TNBbeta distribution can attain three modes when $\varepsilon < 1$. For example, in the symmetric case $p = \nicefrac{1}{2}$, an interior mode exists at $y = \nicefrac{1}{2}$ if and only if $q > \nicefrac{2(1-\varepsilon)}{3}$.  
\end{remark}
 Boundary modes are desirable as they allow for protection against outlier observations and model mis-specification (see~\Cref{sec:robust}). Whereas the beta 
must sacrifice an interior mode to attain boundary modes, the TNBbeta need not. Furthermore, the positive density at the boundaries when $\varepsilon \teq 1$
allows for inclusion of observations $y\teq 0$ and $y\teq 1$ in probability models. Such boundary observations are abundant in practice (see e.g., \Cref{sec:trees}).

In addition to dictating boundary behavior, $\varepsilon$ also impacts the concentration. This is most apparent when $q \teq 0$ and the TNBbeta reduces to a special case of the Libby-Novick beta. 

\newcommand{\corTiltTitle}{\textbf{LNbeta connection}}
\newcommand{\corTiltBody}{%
The TNBbeta density can be written as follows
\begin{equation}
     \TNB(y;\,p,\,q,\,\varepsilon) = \mathrm{LNbeta}\Big(y;\,\varepsilon,\,\varepsilon,\,\tfrac{1\tm p}{p}\Big)(1\tm q)^\varepsilon\big[1-4q\gamma(y,p)\big]^{-(\varepsilon+\nicefrac{1}{2})}.
\end{equation}
Moreover, if $Y \sim \TNB\big(p,\,0,\,\varepsilon\big)$ then, $Y \sim \mathrm{LNbeta}\Big(\varepsilon,\,\varepsilon,\,\frac{1\tm p}{p}\Big)$ as in~\Cref{def:lnb}.
}
\begin{cor}[\corTiltTitle]\label{cor:tilt} \corTiltBody

\end{cor}

\begin{remark}
The similarity function $\gamma(y,p) = \tfrac{1}{4}\mathrm{sech}^2\big(\tfrac{\mathrm{logit}(y)-\mathrm{logit}(p)}{2}\big)$ is the density of the standard logistic distribution centered at $\textrm{logit}(p)$, evaluated at $\textrm{logit}(y)$. This connection stems from \Cref{cor:tilt}, as setting $q\teq 0$ reduces the TNBbeta to $\mathrm{LNbeta}\big(\varepsilon,\varepsilon,\tfrac{1-p}{p}\big)$, whose log-odds is a logistic-beta random variable \citep[e.g.,][]{lee_logistic-beta_2025} centered at $\mathrm{logit}(p)$.
\end{remark}

\begin{remark}\label{remark:alpha}
    As a special case of the TNBbeta, the Libby-Novick beta admits a 
    median-concentration parameterization under equality of its shape 
    parameters, with median $p$ and concentration $\varepsilon$. However, 
    since its concentration $\varepsilon$ enters the beta function, inference on the 
    concentration is intractable. The TNBbeta inherits the median $p$ 
    but introduces $q$ as a tractable concentration parameter, with 
    $\varepsilon$ now only setting a baseline level of concentration.
\end{remark}

Since $\varepsilon$ enters the beta function in the TNBbeta density, we recommend fixing it according to the desired regime of boundary behavior from~\Cref{prop:phi}. We demonstrate in~\Cref{sec:mcmc} that inferences on $p$ and $q$ are tractable in a broad class of regression models.

\subsection{TNBbeta Regression Models}

Given covariates $X_i \in \mathbb{R}^d$ for observations $i \in [n]$, we consider regression models of the form
\begin{equation}\label{eq:pgmodel1.1}
    Y_i \sim \TNB(p_i,\,q_i,\,\varepsilon),\qquad p_i = \textrm{logit}^{-1}\big(X_i^\top \beta_p\big), \qquad q_i = \textrm{logit}^{-1}\big(X_i^\top \beta_q\big)
\end{equation}
in which the parameter $\varepsilon$ is fixed. In particular, the model can accommodate observations of $Y_i \teq 0$ and $Y_i \teq 1$ when $\varepsilon \teq 1$. The coefficient vector $\beta_p$ governs the conditional median of $Y_i$ such that a unit increase in covariate $j$ shifts the log-odds of the median by $(\beta_p)_j$. Similarly, $(\beta_q)_j$ shifts the log-odds of $q_i$, which controls concentration around the median.

This model parallels standard beta regression~\citep{ferrari_beta_2004} and the recently introduced \textit{continuous binomial} (cobin) regression~\citep{lee2026scalable}, both of which connect covariates to the mean. We show in~\Cref{sec:robust} that modeling the median rather than the mean makes TNBbeta regression more robust to outliers than both alternatives.

We demonstrate that inference on $\beta_p$ and $\beta_q$ in~\Cref{eq:pgmodel1.1} is fully conditionally conjugate via a new P\'{o}lya-gamma augmentation for the Libby-Novick beta derived in~\Cref{sec:pglnb} and the latent variable constructions of the TNBbeta distribution developed in~\Cref{sec:mixing}.

\section{Negative Binomial Mixing Distributions}\label{sec:mixing}

We now derive the auxiliary-count construction of the TNBbeta and 
establish conjugate relationships between its negative binomial 
auxiliary variables and the standard beta.

\newcommand{\thmTripleTitle}{\textbf{Triply negative binomial mixing}}
\newcommand{\thmTripleBody}{%
Let $\varepsilon \!>\!0$, $p \!\in\! (0,1)$, $q \!\in\! [0,1]$, and define the following auxiliary-variable randomization of a standard beta random variable
\begin{equation}\bodylabel{eq:triplenb}
    \begin{aligned}
    C \sim \mathrm{NB}\big(\varepsilon,\,1\tm q&\big),\quad A \mid C \!=\! c \sim \mathrm{NB}\big(\varepsilon \!+\! c,\, 1\!-\!p\big),\quad B \mid C \!=\! c\sim \mathrm{NB}\big(\varepsilon \!+\! c, \,p\big),\\
    &Y \mid A \!=\! a, \,B \!=\! b,\, C \!=\! c \sim \mathrm{beta}\big(\varepsilon \!+\! c \!+\! a,\,\, \varepsilon \!+\! c \!+\! b\big).
    \end{aligned}
\end{equation}
Then the marginal distribution is $Y \sim \TNB(p,\,q,\,\varepsilon)$.
}
\begin{theorem}[\thmTripleTitle]\label{thm:triple} \thmTripleBody
\end{theorem}
We prove~\Cref{thm:triple} and all results in this section in~\Cref{app:proof2}.~\Cref{thm:triple} shows that the TNBbeta is a standard beta whose shape parameters depend on 
three negative binomial auxiliary variables---hence 
``triply-randomized.'' This construction simplifies when $q \teq 0$, and thus $C$ is zero, and leads to a new auxiliary-count construction for the Libby-Novick beta. \newcommand{\corLnbNbTitle}{\textbf{Doubly negative binomial mixing}}
\newcommand{\corLnbNbBody}{%
Let $\varepsilon \!>\!0$ and $p \!\in\! (0,1)$. Define the following auxiliary-variable randomization of a standard beta random variable
\begin{equation} \bodylabel{eq:beta_nb}
\begin{aligned}
&A \sim \mathrm{NB}\big(\varepsilon,\,1\tm p\big), \qquad B \sim \mathrm{NB}(\varepsilon,\,p),\\
    &Y \mid A \teq a,\, B \teq b\sim \mathrm{beta}\big(\varepsilon \tp a,\, \varepsilon \tp b\big).
\end{aligned}
\end{equation}
Then the marginal distribution is $Y \sim \mathrm{LNbeta}\Big(\varepsilon,\,\varepsilon,\,\tfrac{1-p}{p}\Big)$.
}\begin{cor}[\corLnbNbTitle]\label{cor:lnb_nb} \corLnbNbBody
\end{cor}

\begin{figure}
\begin{center}
\includegraphics[width=\linewidth]{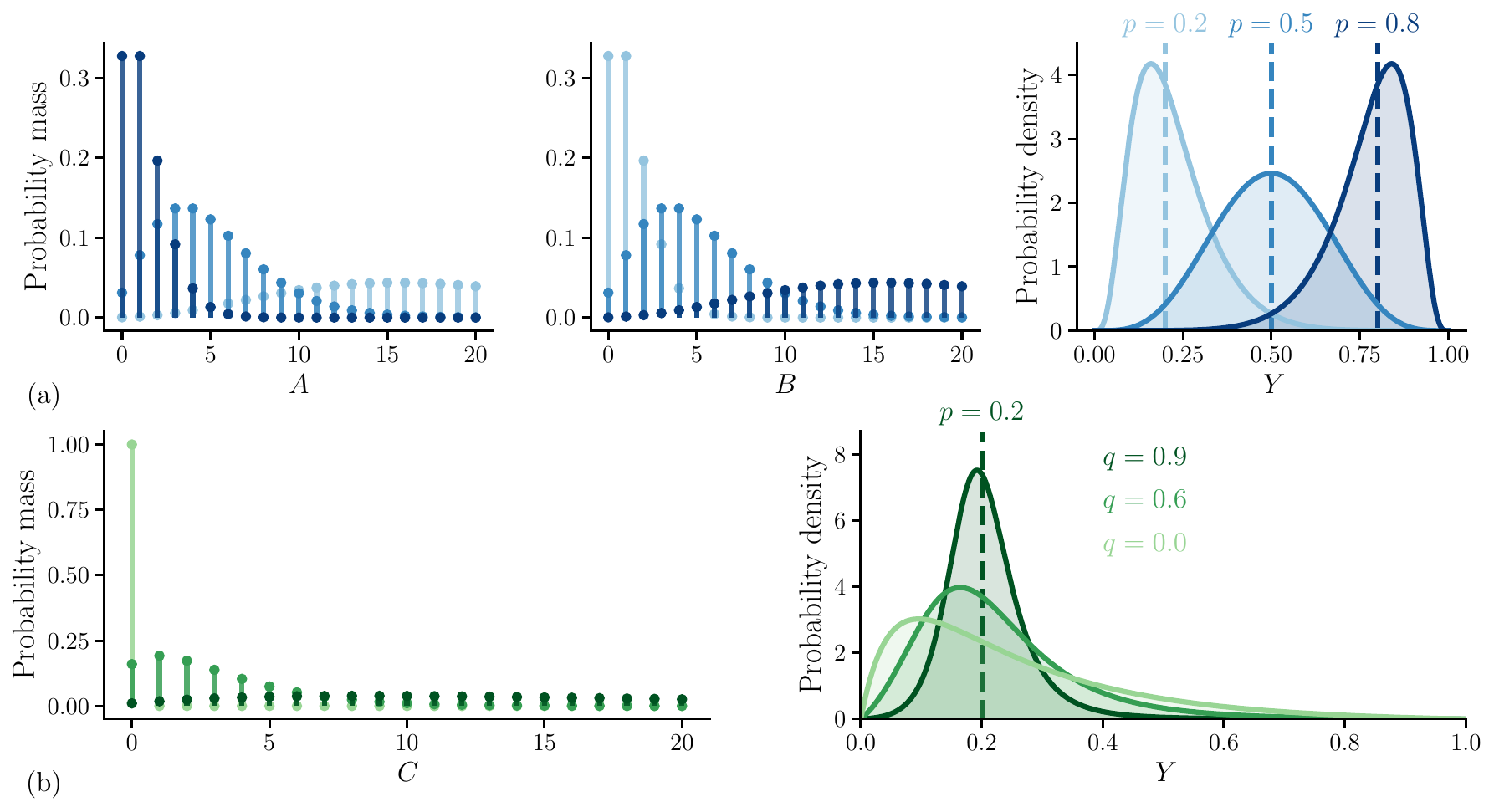}
\end{center}
\caption{(a) A shift in $p$ shifts the relative distribution of probability mass between $A\!\sim\! \mathrm{NB}(5,\,1\tm p)$ and $B\!\sim\! \mathrm{NB}(5,\,p)$ and the induced marginal distribution $Y \!\sim\! \mathrm{LNbeta}\big(5,\,5,\,\tfrac{1-p}{p}\big)$. (b) As $q$ 
increases, $C \!\sim\! \mathrm{NB}(2,\,1\tm q)$ distributes mass to larger values, concentrating the induced marginal distribution $Y \!\sim\! \mathrm{TNBbeta}(0.2,\,q,\,2)$ around its median $p\teq0.2$.\label{fig:Kplot}}
\end{figure}

\Cref{cor:lnb_nb} states that the LNbeta has a bivariate auxiliary-count representation---as in~\Cref{eq:doublecount}---in addition to the single auxiliary version provided by~\cite{chabot_sur_2016} and recalled in~\Cref{eq:nblnb1}. The key to this new bivariate construction is the coupled probability parameters $p$ and $1\tm p$ of the negative binomial auxiliary variables. Intuitively, $A$ and $B$ perturb the conditional distribution of $Y$ away from a shared background distribution $\textrm{beta}(\varepsilon,\,\varepsilon)$. If $A \!>\! B$, the marginal distribution shifts closer to $0$; conversely, if $B \!>\! A$, it shifts closer to $1$, as illustrated in \Cref{fig:Kplot}a. Since the two auxiliary variables share a common shape $\varepsilon$ and differ only through $1\tm p$ and $p$, increasing $p$ increases the perturbation toward $1$. In fact,~\Cref{theorem:tnbebta} shows that this perturbation results in $p$ as the median of $Y$.

\begin{figure}
\begin{center}
\includegraphics[width=\linewidth]{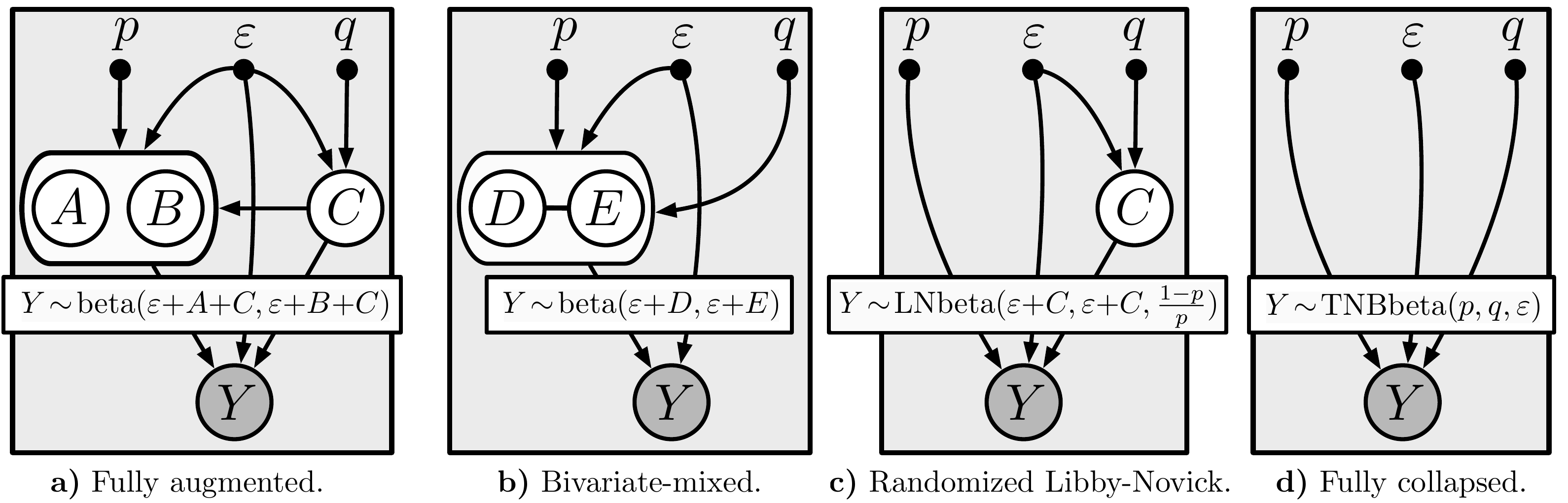}
\end{center}
\caption{\label{fig:gen} TNBbeta generative processes as described in (a)~\Cref{thm:triple}, (b)~\Cref{prop:gen}, (c)~\Cref{cor:lnb_nb}, and (d)~\Cref{prop:ourbetaNB}.}
\end{figure}

The role of the concentration parameter $q$
is clarified by an alternative auxiliary-variable construction of the TNBbeta (see~\Cref{fig:gen}c) which falls out of~\Cref{thm:triple,cor:lnb_nb}.

\newcommand{\corKTitle}{\textbf{Randomized LNbeta}}
\newcommand{\corKBody}{%
Let $\varepsilon \!>\!0$, $p \!\in\! (0,1)$, and $q \!\in\! (0,1)$.  If $\,Y \mid C \teq c \sim \mathrm{LNbeta}\big(\varepsilon + c,\,\varepsilon + c,\,
    \tfrac{1-p}{p}\big)$ and $C \sim \mathrm{NB}\big(\varepsilon,\,1\tm q\big)$, then the marginal is $Y \sim \TNB(p,\,q,\,\varepsilon)$.
}\begin{cor}[\corKTitle]\label{cor:K} \corKBody
\end{cor}

Since the LNbeta concentration increases with its shape parameters, $C$ adds concentration beyond the baseline
set by $\varepsilon$, with larger $q$  
concentrating $Y$ around its median (\Cref{fig:Kplot}b).

Note that $C$ performs a dual role in the fully augmented construction in~\Cref{fig:gen}a. First, $A$ and $B$ share not only a dependence on $p$ via coupled probability parameters but also a dependence on $C$ through their shapes. Second, the standard beta is randomized with the counts $D \teq A \tp C$ and $E \teq B \tp C$. Both roles of $C$ induce dependence on the bivariate mixing distribution $(D,E) \sim \mathcal{DM}$ illustrated in~\Cref{fig:gen}b and which we characterize below.

\newcommand{\propGenTitle}{\textbf{Mixing distribution}}
\newcommand{\propGenBody}{%
Define $D = A \tp C$ and $E = B \tp C$ where
\begin{equation}
            C \sim \mathrm{NB}\big(\varepsilon,\,1 \tm q\big),\qquad A \mid C = c \sim \mathrm{NB}\big(\varepsilon \tp c,\, 1\tm p\big),\qquad B \mid C = c\sim \mathrm{NB}\big(\varepsilon \tp c,\, p\big).
    \end{equation}
    Then the discrete bivariate distribution $\mathcal{DM}$ for $(D,E)$ has probability mass function
    \begin{equation}\bodylabel{eq:m1m2}
        P(d,\,e;\,p,\,q,\,\varepsilon) = \mathrm{NB}(d;\,\varepsilon,\,1\tm p)\cdot\mathrm{NB}(e;\,\varepsilon,\,p)\cdot(1 \tm q)^\varepsilon\;{}_2F_1(-d,-e;\,\varepsilon;\,q)
    \end{equation}
    where ${}_2F_1(-d,-e;\,\varepsilon;\,q)$ is a terminating Gauss hypergeometric series.
}
\begin{proposition}[\propGenTitle]\label{prop:gen} \propGenBody
\end{proposition}
With $(D,\,E) \sim \mathcal{DM}$ as in~\Cref{eq:m1m2}, the auxiliary-count representation of the TNBbeta can be alternatively expressed as $Y \!\mid\! D \teq d,\,E \teq e \!\sim\! \textrm{beta}(\varepsilon + d,\,\varepsilon + e)$. The results thus far show that this particular choice of mixing distribution for the beta leads to appealing analytic properties, including a closed-form marginal distribution for $Y$ parameterized by its median $p$ and concentration $q$. Although we do not use~\Cref{eq:m1m2} directly for parameter inference, we show below that this choice of mixing distribution induces analytic properties for efficient parameter inference of both $p$ and $q$ in a variety of hierarchical models.

We show that leveraging the auxiliary-count representations in~\Cref{eq:triplenb,eq:beta_nb} is key for computational tractability in such models. First, when $q\teq 0$ and $Y$ is thus Libby-Novick beta-distributed, as in~\Cref{cor:lnb_nb}, we show that the inverse distribution $(A,\,B)\mid Y$ has a simple analytic form. In fact, the relationship is conjugate.

\newcommand{\propConTitle}{\textbf{Inverse distributions for $A$ and $B$}}
\newcommand{\propConBody}{%
Define $A$, $B$, and $Y$ as in~\Cref{cor:lnb_nb}. Then the inverse distribution for $A$ is negative binomial
    \begin{equation}\bodylabel{eq:conjugacy}
        A \mid Y \teq y \sim \mathrm{NB}\Big(2\varepsilon,\,\, 1 \tm \tfrac{py}{1 \tm (1\tm p)(1 \tm y)}\Big)
    \end{equation}
    Moreover, the distribution of $A$ conditioned on both $Y$ and $B$ is also negative binomial
    \begin{equation}\bodylabel{eq:conditionalconjugacy}
        A \mid B \teq b,\, Y \teq y \sim \mathrm{NB}\big(2\varepsilon \tp b,\, 1 \tm py\big)
    \end{equation}
    In each case, the distribution for $B$ is identical up to substitution of $py$ with $(1\tm p)(1\tm y)$. Thus the conditional distribution $(A,B) \mid Y$ is a product of negative binomial densities.
}
\begin{proposition}[\propConTitle]\label{prop:con} \propConBody
\end{proposition}
It is well-known that for a negative binomial likelihood $\textrm{NB}(y;\,r,\,q)$ with success parameter $q$, the conjugate prior for $q$ is beta---i.e., $q \mid Y$ is beta-distributed when $q \sim \textrm{beta}(\cdot,\cdot)$.~\Cref{prop:con} establishes a form of conjugacy in the reverse direction, in which the negative binomial is the conjugate prior for a beta likelihood. Furthermore, each negative binomial is conditionally conjugate given the other, due to~\Cref{eq:conditionalconjugacy}. Moreover, conditional distributions involving the sum $A \tp B$ are also remarkably simple, despite the fact that the marginal distribution for $A \tp B$ is not in general negative binomial.
\newcommand{\propConsumTitle}{\textbf{Inverse distribution for $A\tp B$}}
\newcommand{\propConsumBody}{ Define $A$, $B$, and  $Y$ as in~\Cref{cor:lnb_nb}. Then the inverse distribution for the sum $A + B$ is also negative binomial
\begin{equation}
    A \tp B\mid Y \teq y \sim \mathrm{NB}\big(2\varepsilon,\,\, y(1-p) + p(1-y)\big).
\end{equation}
Moreover, the distribution for $A$ conditional on the sum $A \tp B$ and $Y$ is binomial
\begin{equation}
    A \mid A \tp  B \teq j,\, Y \teq y \sim \mathrm{binomial}\Big(j, \,\mathrm{logit}^{-1}\big(\logit(y) + \logit(p)\big)\Big).
\end{equation}
}
\begin{proposition}[\propConsumTitle]\label{prop:consum} \propConsumBody
\end{proposition}
\Cref{prop:con,prop:consum} each provide strategies for sampling 
$(A,\,B) \mid Y,\,C$ which we exploit in~\Cref{sec:mcmc} for inference 
in TNBbeta regression models. We show that these results are especially useful when $\varepsilon \teq 1$ and boundary values $Y \teq 0$ and $Y \teq 1$ are observed. This approach first requires sampling from $C \mid Y$,
which we show is also negative binomial.

\newcommand{\propKconditionalTitle}{\textbf{Inverse distribution for $C$}}
\newcommand{\propKconditionalBody}{%
The inverse distribution of the negative binomial $C$ in the auxiliary-count representation in~\Cref{eq:triplenb} is negative binomial
    \begin{equation}\bodylabel{eq:Kconditional}
        C \mid Y = y\sim \text{NB}\big(\varepsilon \tp \tfrac{1}{2},\,1 \tm 4q\gamma(y,p)\big)
    \end{equation}
    where the similarity function $\gamma(y,p)$ is defined in~\Cref{eq:g}.
}
\begin{proposition}[\propKconditionalTitle]\label{prop:Kconditional} \propKconditionalBody
\end{proposition}

We exploit~\Cref{prop:Kconditional} to construct auxiliary-variable Gibbs samplers to infer $p$ and $q$ in models which assume the TNBbeta as a likelihood. Intuitively, if $C$ were known, inference on $p$ reduces to as if $Y$ were Libby-Novick beta, due to~\Cref{cor:lnb_nb}. Moreover, inference on $q$ reduces to that for the negative binomial $C$. Thus, using properties of negative binomial and Libby-Novick beta random variables, we show that a ``data-augmentation'' approach for parameter inference which instantiates $C$ leads to tractable inference via conditional conjugacy in many hierarchical models, including TNBbeta regression models of the form
\begin{equation}\label{eq:pgmodel2}
    Y_i \sim \TNB(p_i,\,q_i,\,\varepsilon),\qquad p_i = \textrm{logit}^{-1}\big(X_i^\top \beta_p\big), \qquad q_i = \textrm{logit}^{-1}\big(X_i^\top \beta_q\big).
\end{equation}
Moreover, we show that instantiating the latent variables $A$ and $B$ will often not be necessary for inference in these models. In addition to~\Cref{prop:Kconditional}, the key ingredient for inference is a new Pólya-gamma augmentation scheme for $(0,\,1)$-bounded support data. We develop the necessary tools for posterior inference for such models in~\Cref{sec:mcmc}, emphasizing the essential role of the auxiliary-count representations described in this section.

\section{Data Augmentation for TNBbeta Regression Models}\label{sec:mcmc}
This section develops a data-augmentation framework for inference in TNBbeta models. Such models assume $Y_i \sim \TNB(p_i,\,q_i,\,\varepsilon)$ where data share a fixed $\varepsilon$ and have potentially different median $p_i$ and concentration $q_i$, each of which may be functions of global parameters. 

Our approach to inference centers around the auxiliary count representation in~\Cref{thm:triple} and the use of data augmentation~\citep{tanner_calculation_1987} to instantiate these counts as latent variables. The key to this approach is the inverse distribution of~\Cref{prop:Kconditional}. Each data point $Y_i \sim \TNB(p_i,\,q_i,\,\varepsilon)$ can be alternatively expressed via~\Cref{cor:K} as\begin{equation}
     C_i \sim \text{NB}\big(\varepsilon,\,1\tm q_i\big),\,\,\,\, Y_i \mid C_i \teq c_i,\, p_i \sim  \mathrm{LNbeta}\big(\varepsilon + c_i,\,\varepsilon + c_i,\,
    \tfrac{1-p_i}{p_i}\big).
\end{equation}
Performing the computationally cheap and conceptually simple data augmentation step,
\begin{equation}\label{eq:dataK}
    C_i \mid Y_i \teq y_i,\,p_i,\,q_i\sim \text{NB}\big(\varepsilon + \tfrac{1}{2},\,\,1 - 4q_i\gamma(y_i,\,p_i)\big),
\end{equation}
greatly simplifies inference on $p_i$ and $q_i$. Conditional on $C_i$, inference on $q_i$ depends only on $C_i$---and not on $Y_i$---so any technique for negative binomial probability parameters applies directly. If $q_i \sim \textrm{beta}(\cdot,\cdot)$, for example, then $q_i \mid C_i$ is also beta by conjugacy.

Inference on $p_i$ involves both $C_i$ and $Y_i$. Conditional on $C_i$, each $Y_i$ is Libby-Novick beta-distributed, so a second layer of data augmentation---instantiating $A_i$ and $B_i$ via the inverse distributions in~\Cref{prop:con} and updating $p_i \mid C_i, A_i, B_i$---yields a conjugate beta update when $p_i$ is beta-distributed. However, this second layer is often unnecessary. We show below that for regression models, inference on $p_i$ bypasses $A_i$ and $B_i$ for all non-boundary $Y_i \!\in\! (0,1)$ via Pólya-gamma augmentation applied directly to the Libby-Novick beta density.

\subsection{Pólya-gamma Augmentation}\label{sec:pg}

Many probability density functions can be expressed in a ``logistic form'' as follows
\begin{equation}
\label{eq:likelihood}
P(y \mid \psi)
= c(y)\,
\frac{(e^{\psi})^{a(y)}}{(1+e^{\psi})^{b(y)}},
\end{equation}
for known functions $a(y)$, $b(y)$, and $c(y)$. The key to Pólya-gamma augmentation is to re-express a likelihood of the form in~\Cref{eq:likelihood} using the following integral identity,
\begin{equation}
\label{eq:pg_identity}
\frac{(e^{\psi})^{a(y)}}{(1+e^{\psi})^{b(y)}}
= 2^{-b(y)} e^{\kappa(y) \psi}
\int_{0}^{\infty}
e^{-\omega \psi^{2}/2}
\, \mathrm{PG}\big(\omega;\;b(y),0\big)\, d\omega,
\end{equation}
where $\kappa(y) = a(y) - \nicefrac{b(y)}{2}$ and $\mathrm{PG}\big(\omega;\;b(y),0\big)$ is the density of a \textit{Pólya-gamma}
random variable (see~\cite{polson_bayesian_2013} for details on the Pólya-gamma distribution). This identity expresses the density in~\Cref{eq:likelihood} as a Gaussian mixture over $\omega$. Treating $\omega$ as a latent variable yields simple conditional distributions. Conditioned on $\omega$, the complete conditional distribution for $\psi$ is $P(\psi \mid \omega, y)
\propto P(\psi)\, e^{\kappa(y)\psi} e^{-\omega \psi^{2}/2}$
which is Gaussian if the prior $P(\psi)$ is Gaussian. Conversely, by the exponential tilting property of the Pólya-gamma distribution, the conditional distribution for $\omega$ is also Pólya-gamma as $\omega \mid \psi, y \sim \mathrm{PG}\big(b(y), \psi\big)$.

Together, these conditional distributions form a conjugate augmentation scheme for models with Gaussian priors and logistic-type likelihoods, enabling efficient Gibbs sampling and other inference procedures. Pólya-gamma augmentation is most common as a posterior computation technique in models of integer-valued count data, such as in Bernoulli and binomial~\citep{polson_bayesian_2013}, negative binomial~\citep{polson_bayesian_2013,zhou_lognormal_2014}, and multinomial regression models~\citep{polson_bayesian_2013,chen_scalable_2013,linderman_dependent_2015}. Less common is the use of this augmentation in models with a continuous response. The rest of this section extends these results to models with LNbeta and TNBbeta likelihoods.

\subsection{A New Pólya-gamma Augmentation Scheme for the LNbeta}\label{sec:pglnb}

We show that the Libby-Novick beta density is amenable to Pólya-gamma augmentation when the median is parameterized as $p = \textrm{logit}^{-1}(\psi)$. The density is proportional in $\psi$ to
\begin{align}
\mathrm{LNbeta}\big(\varepsilon,\,\varepsilon,\,
    \tfrac{1-p}{p}\big)
&\propto_\psi 
\left(\frac{1}{1 + e^\psi}\right)^\varepsilon \left(\frac{e^\psi}{1 + e^\psi}\right)^\varepsilon \left(\frac{1 + e^\psi}{y + (1-y)e^\psi}\right)^{2\varepsilon}\\
&\propto_\psi  \frac{\big(e^{\psi})^\varepsilon}{(y+(1-y)e^{\psi})^{2\varepsilon}}\\
&\propto_\psi  \frac{(e^{\psi +\mathrm{logit}(1-y)}\big)^\varepsilon}{\big(1+e^{\psi +\mathrm{logit}(1-y)}\big)^{2\varepsilon}}
\end{align}
which is exactly the logistic form of~\Cref{eq:likelihood} with $a(y)=\varepsilon$, $b(y) = 2\varepsilon$, and $\psi +\mathrm{logit}(1-y)$ in place of $\psi$.
Augmenting with $\omega \sim \textrm{PG}(2\varepsilon,\,0)$ yields the conditional distributions
\begin{align}\label{eq:lnbpsi}
       &\psi \mid \omega, \,y
\propto P(\psi)\,e^{-\omega(\psi^2/2 \,+\, \textrm{logit}(1-y)\,\psi)}\\
&\omega \mid \psi,\,y\sim\textrm{PG}\big(2\varepsilon,\, \psi + \textrm{logit}(1-y)\big).
\end{align}
Because $a(y) = \nicefrac{b(y)}{2}$, the term $\kappa(y)$ from the derivations in~\Cref{sec:pg} vanishes, and both conditional distributions depend on the data only through the log-odds offset $\textrm{logit}(1-y)$.

Pólya-gamma augmentation is even possible as an inference strategy for $p$ because it does not enter any special function in its density, unlike the parameters of the beta and most of its other generalizations. As discussed in~\Cref{remark:alpha}, however, the LNbeta concentration $\varepsilon$ enters the beta function and inherits the computational intractability of learning beta shape parameters. In the next section, we build on this result to derive a fully conjugate Gibbs sampler for inference on both median $p$ and concentration $q$ in TNBbeta regression models.

\subsection{Auxiliary-Variable Gibbs Sampling for TNBbeta Regression}
\label{sec:gibbs}

We now present a conditionally conjugate Gibbs sampler for regression models of bounded-support responses $Y_i \in [0,1]$ with covariate profiles $X_i \in \mathbb{R}^d$. Consider the TNBbeta regression model with logit link functions for the median and concentration as follows
\begin{equation}
\begin{aligned}\label{eq:auxmodel}
       Y_i &\sim \TNB(p_i,\,q_i,\,\varepsilon)\\
   p_i = \textrm{logit}^{-1}(&X_i^\top \beta_p)\qquad
   q_i = \textrm{logit}^{-1}(X_i^\top  \beta_q),
\end{aligned}
\end{equation}
for each observation $i \in [n]$. We complete the model with Gaussian priors  $\beta_p \sim \textrm{MVN}(\mathbf{b}_p,\,B_p)$ and $\beta_q \sim \textrm{MVN}(\mathbf{b}_q,\,B_q)$ on the coefficients for the median and concentration, respectively.

Augmenting the model with one negative binomial and two Pólya-gamma random variables per data point yields Gaussian updates for both $\beta_p$ and $\beta_q$. As described above, the first step is to 
introduce the negative binomial random variable $C_i \sim \mathrm{NB}\big(\varepsilon,\,1\tm q_i\big)$ to the model via the alternative representation of the TNBbeta described in~\Cref{cor:K}. Instantiating $C_i$ during inference requires sampling it from its complete conditional distribution as follows\begin{equation}\label{eq:nbaug}
C_i \mid Y_i \teq y_i,\,p_i,\,q_i\sim \text{NB}\big(\varepsilon + \tfrac{1}{2},1 \tm 4q_i\gamma(y_i,p_i)\big),
\end{equation}
as shown in~\Cref{prop:Kconditional}. Once 
$C_i$ is instantiated for each observation, inference on the coefficient vectors is greatly simplified. As $\beta_q$ is conditionally independent of the data $\{Y_i\}_{i=1}^n$ given $\{C_i\}_{i=1}^n$, inference on $\beta_q$ can proceed through $\{C_i\}_{i=1}^n$ alone and follows from the standard Pólya-gamma augmentation scheme for the negative binomial~\citep{polson_bayesian_2013}. Augmenting the model with Pólya-gamma random variables $\omega_i \sim \mathrm{PG}(\varepsilon \tp c_i,\,0)$ then results in a Pólya-gamma complete conditional for each $\omega_i$ and a Gaussian update for $\beta_q$ as
\begin{align}
    &\omega_i \mid C_i \teq c_i,\, \beta_q \sim \textrm{PG}\big(\varepsilon + c_i,\, X_i^\top \beta_q\big)\\
    &\beta_q \mid \{\omega_i,\,C_i\}_{i=1}^n \sim  \textrm{MVN}\big(V_q (X^\top \boldsymbol{\kappa} \tp B_q^{-1}\boldsymbol{b}_q),\, V_q \big) 
\end{align}
in which we define the covariance matrix $V_q = (X^\top \Omega X  + B_q^{-1})^{-1}$ with $\Omega = \textrm{diag}(\boldsymbol{\omega})$ as a diagonal matrix of the auxiliary Pólya-gamma variables and $\boldsymbol{\kappa}$ as the vector of $\kappa_i = \nicefrac{\varepsilon - C_i}{2}$. 

Introducing $C_i$ also enables Pólya-gamma augmentation for inference of $\beta_p$. Conditional on $C_i$, each $Y_i$ is Libby-Novick beta-distributed, so we can apply the Pólya-gamma augmentation derived in~\Cref{sec:pglnb}. Therefore, introducing $\xi_i \sim \mathrm{PG}\big(2(\varepsilon + C_i),\,0\big)$ yields the updates
\begin{align}
    &\xi_i \mid C_i \teq c_i,\, Y_i \teq y_i,\, \beta_p\, \sim \textrm{PG}\big(2(\varepsilon \tp c_i),\;X_i^\top \beta_p + \textrm{logit}(1\tm y_i)\big)\label{eq:startpolya}\\
    &\beta_p \mid \{\xi_i,\,Y_i\}_{i=1}^n \sim \textrm{MVN}\big(V_p(B_p^{-1} \boldsymbol{b}_p  \tm X^\top \boldsymbol{\eta}),\,V_p \Big)\label{eq:endaug}
\end{align}
where we define the analogous covariance matrix $V_p = (X^\top \Xi X + B_p^{-1})^{-1}$ with the diagonal matrix of Pólya-gamma auxiliary variables $\Xi \teq \textrm{diag}(\boldsymbol{\xi})$ and further define $\boldsymbol{\eta} \teq \boldsymbol{\xi} \odot \logit(1 -\boldsymbol{y})$ where $\odot$ is the elementwise product and $\logit$ is applied elementwise to the data vector $\boldsymbol{y}$.

Together,~\Crefrange{eq:nbaug}{eq:endaug} constitute a Gibbs sampler for inference on the coefficient vectors $\beta_p$ and $\beta_q$. This algorithm relies only on the ability to sample multivariate normal, negative binomial, and Pólya-gamma random variables. Due to the ubiquity of Pólya-gamma augmentation as a tool for posterior computation in models of discrete data, efficient algorithms for Pólya-gamma sampling are now available in most statistical software.

Recall that TNBbeta models can accommodate exact boundary observations $Y_i \teq 0$ or $Y_i \teq 1$ when $\varepsilon\teq 1$ and the TNBbeta likelihood admits positive, finite density at the boundary. However, the complete conditional for each Pólya-gamma variable $\xi_i$ depends on  $Y_i$ via the log-odds offset $\textrm{logit}(1-Y_i)$ and therefore cannot handle exact observations of zero or one. 

For boundary observations, we introduce additional auxiliary variables $A_i \sim \mathrm{NB}\big(\varepsilon,\,1\tm p_i\big)$ and $B_i \sim \mathrm{NB}(\varepsilon,\,p_i)$ into the model as in the TNBbeta representation of~\Cref{thm:triple}, where $Y_i \teq 0$ or $Y_i \teq 1$ implies $C_i \teq 0$.
Then by~\Cref{prop:con}, $A_i$ and $B_i$ can be instantiated via
\begin{equation}
(A_i, B_i) \mid Y_i \teq y_i,\, p_i \sim
{\renewcommand{\arraystretch}{0.7}\begin{cases}
\big(\delta_0,\; \text{NB}(2\varepsilon,\, p_i)\big) & \!\!\textrm{if } y_i = 0 \\
\big(\text{NB}(2\varepsilon,\, 1\tm p_i),\; \delta_0\big) & \!\!\textrm{if } y_i = 1
\end{cases}}
\end{equation}
Then using again Pólya-gamma augmentation for negative binomial random variables, we are able to instantiate each $\xi_i$ according to its complete conditional distribution
\begin{equation}
    \xi_i \mid A_i \teq a_i,\, B_i \teq b_i,\, \beta_p \sim \textrm{PG}\big(2\varepsilon \tp a_i \tp b_i,\;X_i^\top \beta_p\big).
\end{equation}
Conditional on the auxiliary variables $\{\xi_i\}_{i=1}^n$ for all observations, $\beta_p$ is drawn as in~\Cref{eq:endaug} where now $\eta_i \teq \nicefrac{b_i}{2}$ when $Y_i \teq 0$ and $\eta_i \teq -\nicefrac{a_i}{2}$ when $Y_i \teq 1$. These modifications to the Gibbs sampler enable inference of $\beta_p$ in the presence of boundary observations.

The model in~\Cref{eq:auxmodel} is one of many for which this framework yields tractable posterior inference. For example, if the concentration is a shared parameter $q \sim \textrm{beta}(\cdot,\cdot)$, the Gibbs sampler further simplifies, as the complete conditional for $q$ is conjugate beta.

More broadly, because the complete conditional distributions for $\beta_p$ and $\beta_q$ are multivariate normal, this framework inherits the computational toolkit of Gaussian linear models, including spatial, temporal, and hierarchical priors. We demonstrate these capabilities in the next section, where we fit several variants of this model to $(0,1)$-bounded support data.

\section{Experimental Results}\label{sec:experiments}

We now compare the performance of TNBbeta regression models to alternative bounded-support regression models in 
both synthetic and real-data scenarios, demonstrating that TNBbeta regression models are more robust to noise, often achieve better held-out prediction, and naturally incorporate complex prior structure such as spatial random effects.

\subsection{Predictive Evaluation}
\label{sec:eval_metrics}

To assess model fit in each experiment, we hold out a portion of the data $\mathbf{Y}_\textrm{held} \equiv \{Y_i\}_{i \in n_\textrm{held}}$ from the observed data $\mathbf{Y}_\textrm{obs} \equiv \{Y_i\}_{i \in n_\textrm{obs}}$, where $n_\textrm{obs}$ and $n_\textrm{held}$ partition the data indices.  Model fit is assessed via \textit{log pointwise predictive density} (LPPD) of the held-out data~\citep{gelman_understanding_2014}, defined as $\textrm{LPPD} = \sum_{i \in n_\textrm{held}}\log P(Y_i \mid \mathbf{Y}_\textrm{obs})$. We report a function of LPPD called the \textit{information rate (IR)}, which measures the average surprise (i.e., Shannon information) of new observations:
\begin{align}\label{eq:inforate}
\textrm{IR} = -\tfrac{1}{|n_\textrm{held}|
}\underbrace{\sum_{i \in n_\textrm{held}}\log P(Y_i \mid \mathbf{Y}_\textrm{obs})}_{=\textrm{LPPD}} &= -\tfrac{1}{|n_\textrm{held}|
}\sum_{i \in n_\textrm{held}} \log \int f_{\theta_i}(Y_i)\, P(\theta_i \mid \, \mathbf{Y}_\textrm{obs}) \mathbf{d}\theta_i
\end{align}
where $f_{\theta_i}$ denotes the model density with parameters $\theta_i$ that may depend on global parameters $\Theta$. This defines a posterior expectation which we approximate with Monte Carlo as $\textrm{IR} \approx -\tfrac{1}{|n_\textrm{held}|}\sum_{i \in n_\textrm{held}} \log \left[\tfrac{1}{S} \sum_{s=1}^S f_{\theta_i^{\ms{(s)}}}(Y_i)\right]$ using the $S$ posterior samples $\theta_i^{\ms{(s)}} \sim P(\theta_i \mid \, \mathbf{Y}_\textrm{obs})$ returned by MCMC. Lower information rate indicates better predictive fit. To facilitate direct comparison for a TNBbeta model against a beta model baseline, we also report \textit{information gain (IG)} defined as $\text{IG} = \textrm{IR}_{\textrm{baseline}} - \textrm{IR}$, which measures the reduction in average surprise from the baseline, where higher is better.

\begin{figure}[ht]
\begin{center}
\includegraphics[width=\linewidth]{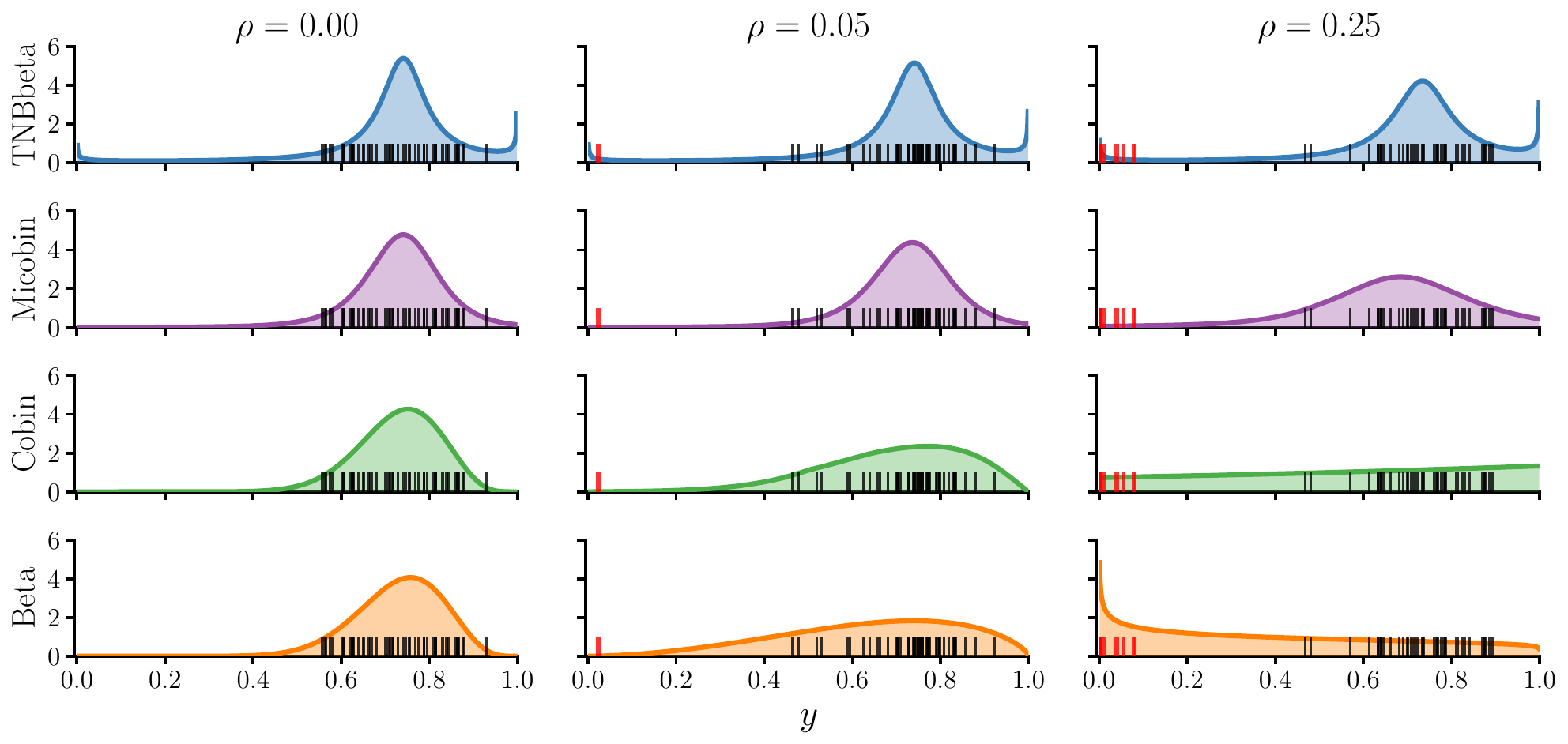}
\end{center}
\caption{The posterior predictive for TNBbeta models is much less sensitive to increasing levels of noise than that for alternative models, as shown for $G_1$ observations. \label{fig:postpred}}
\end{figure}

\subsection{Robustness to Noise}
\label{sec:robust}

In this section we compare the performance under noise of TNBbeta regression to that of existing alternatives. Its median parameterization and boundary modes (for $\varepsilon < 1$) suggest that the TNBbeta should be 
less sensitive to outliers than mean-parameterized 
models.

\begin{figure}
\begin{center}
\includegraphics[width=\linewidth]{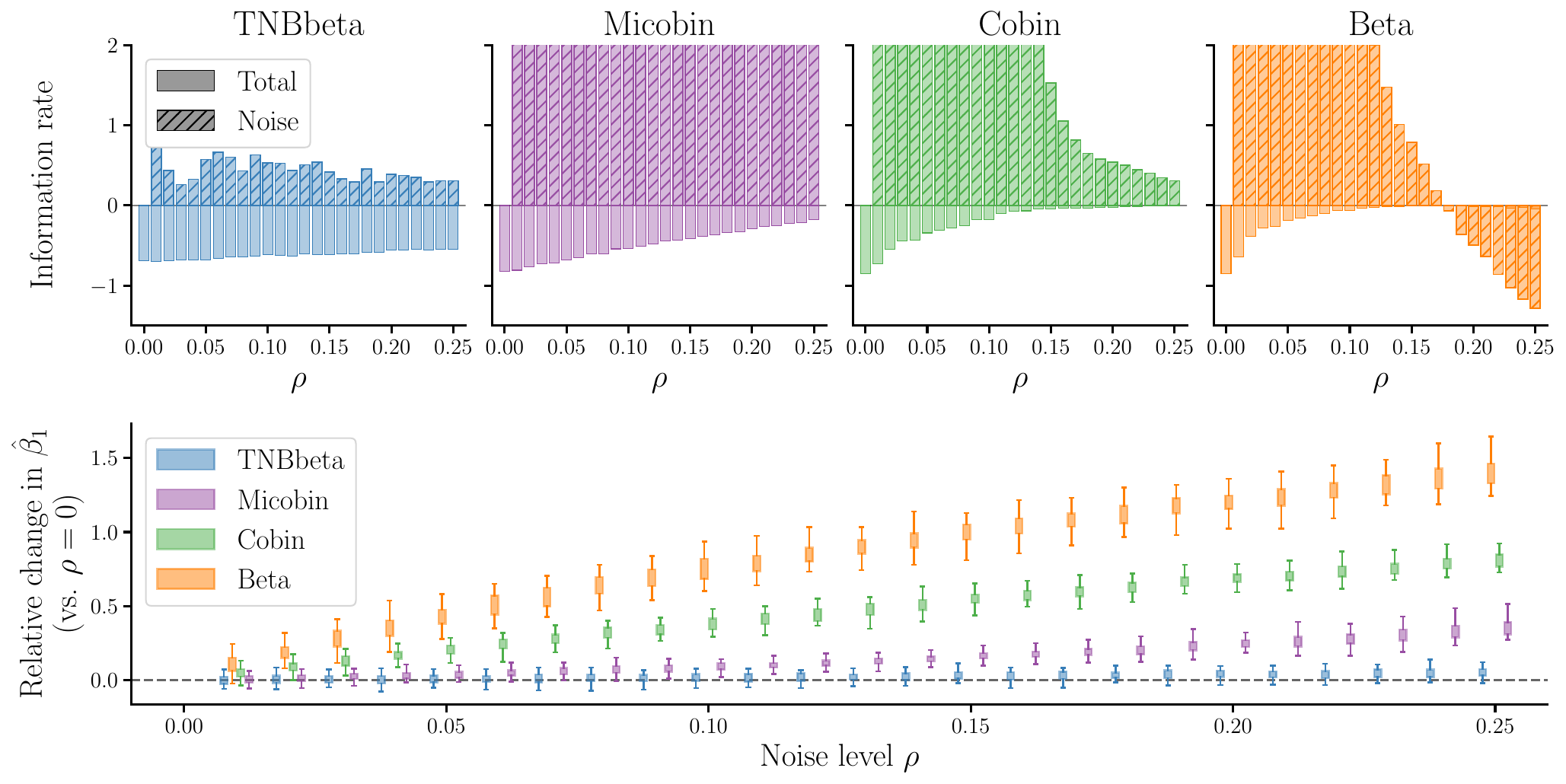}
\end{center}
\caption{TNBbeta regression models are far more robust to noise than beta, cobin, and micobin alternatives in terms of (a) information rate $(\downarrow)$ on all held-out data and all noise observations and (b) difference in $\beta_1$ estimates compared to the estimate without noise $(\downarrow)$. \label{fig:robust}}
\end{figure}

\textbf{Data.} We generate $N=1000$ observations $Y_i \sim \textrm{beta}\big(\phi\mu_i,\,\phi(1-\mu_i)\big)$ with $\mu_i = \textrm{logit}^{-1}\big(\beta_0^* + \beta_1^* X_i\big)$, where $\beta_0^*=0$, $\beta_1^*=1$, and $\phi=20$. Observations are split into equal groups $G_0$ ($X_i=0$) and $G_1$ ($X_i=1$). In $G_1$, a fraction $\rho\in[0,0.25]$ of observations are replaced by noise drawn from $\textrm{beta}\big(\phi\mu_\rho,\,\phi(1-\mu_\rho)\big)$ with $\mu_\rho = \textrm{logit}^{-1}\big(\beta_0^* - 4\beta_1^*\big)$.

\textbf{Models.} We fit four regression models to each dataset. The TNBbeta model specifies $Y_i \sim \TNB\big(p_i,\,q,\,\varepsilon\big)$ with median $p_i = \textrm{logit}^{-1}\big(\beta_0 + \beta_1 X_i\big)$, $q \sim \textrm{beta}(1,1)$, and $\varepsilon = 0.5$. We also fit beta regression models $Y_i \sim \textrm{beta}\big(\phi \mu_i,\,\phi(1-\mu_i)\big)$ with mean $\mu_i = \textrm{logit}^{-1}\big(\beta_0 + \beta_1 X_i\big)$ and $\phi \sim \textrm{lognormal}(0,1)$. We also compare to the continuous binomial (cobin) and mixture of continuous binomials (micobin) models of~\cite{lee2026scalable}, which use a specialized \textit{cobin} link function to connect $\beta_0 + \beta_1 X_i$ to the mean. In each case, we specify $\beta_0, \beta_1 \sim \mathcal{N}(0,1)$.

\textbf{Implementation details.}  For each noise level, we generate $50$ datasets and fit each with $4$ MCMC chains, collecting $500$ samples after $500$ iterations of burn-in.

\textbf{Results.} \Cref{fig:postpred} shows the posterior predictive for $G_1$ observations for each model at three noise levels $\rho \in \{0,.05,0.25\}$. While each posterior predictive for the alternative models shifts to accommodate noise observations, the TNBbeta posterior predictive is largely unchanged. Moreover, with the choice of $\varepsilon \teq 0.5$, the mode at $y\teq 0$ allows the TNBbeta model to absorb boundary outliers without distorting predictions near the median. The benefit of the boundary mode can also be seen in~\Cref{fig:robust}a. The TNBbeta fits far better to noise observations for most values of $\rho$, and its predictions are stable across $\rho$. In contrast, the other models initially fit very poorly to outlying observations but quickly overfit to them, sacrificing overall predictive performance. In addition to prediction, TNBbeta estimation is also far more stable, as can be seen in~\Cref{fig:robust}b which shows largely unchanged estimates of $\beta_1$ under increasing levels of noise. Overall, the TNBbeta model is far more robust to noise in this setting, even when the underlying data are truly beta-distributed.

\begin{table}[ht]
\centering
\caption{TNBbeta regression models often attain lower information rate than alternatives and always produce more effective samples per second than data augmentation alternatives (cobin and micobin), though HMC-based beta regression can be more efficient.}
\label{tab:bench_single}

\small
\setlength{\tabcolsep}{3pt}
\begin{tabular}{l cc cc cc cc cc}
\toprule
 & &  & \multicolumn{2}{c}{TNBbeta} & \multicolumn{2}{c}{Beta} & \multicolumn{2}{c}{Cobin} & \multicolumn{2}{c}{Micobin} \\
 \cmidrule(lr){4-5} \cmidrule(lr){6-7} \cmidrule(lr){8-9} \cmidrule(lr){10-11}
Dataset & $n$ & $p$ & IR $(\downarrow)$ & $\nicefrac{\textrm{ESS}}{\textrm{sec}}$ & IR $(\downarrow)$ & $\nicefrac{\textrm{ESS}}{\textrm{sec}}$ & IR $(\downarrow)$ & $\nicefrac{\textrm{ESS}}{\textrm{sec}}$ & IR $(\downarrow)$ & $\nicefrac{\textrm{ESS}}{\textrm{sec}}$ \\
\midrule
Orthogonal & 500 & 20 & -0.35 \scriptsize{(0.02)} & 368 & \textbf{-0.42 \scriptsize{(0.02)}} & \textbf{1260} & -0.32 \scriptsize{(0.03)} & 215 & -0.35 \scriptsize{(0.02)} & 67 \\[2pt]
Collinear & 500 & 20 & \textbf{-0.31 \scriptsize{(0.02)}} & \textbf{353} & \textbf{-0.33 \scriptsize{(0.02)}} & 174 & \textbf{-0.31 \scriptsize{(0.02)}} & 259 & \textbf{-0.31 \scriptsize{(0.02)}} & 74 \\[2pt]
Bike & 731 & 11 & \textbf{-1.61 \scriptsize{(0.02)}} & \textbf{266} & \textbf{-1.59 \scriptsize{(0.02)}} & 64 & -1.46 \scriptsize{(0.02)} & 41 & -1.49 \scriptsize{(0.02)} & 12 \\[2pt]
Product & 1,160 & 10 & \textbf{-0.67 \scriptsize{(0.01)}} & 158 & -0.57 \scriptsize{(0.01)} & \textbf{236} & -0.55 \scriptsize{(0.01)} & 93 & -0.64 \scriptsize{(0.02)} & 19 \\[2pt]
Alcohol & 1,288 & 3 & -2.18 \scriptsize{(0.02)} & 100 & \textbf{-2.30 \scriptsize{(0.02)}} & \textbf{296} & -2.26 \scriptsize{(0.01)} & 10 & -2.27 \scriptsize{(0.01)} & 4 \\[2pt]
Crime & 1,940 & 20 & \textbf{-0.99 \scriptsize{(0.02)}} & \textbf{77} & -0.88 \scriptsize{(0.02)} & 14 & -0.94 \scriptsize{(0.02)} & 49 & -0.93 \scriptsize{(0.02)} & 13 \\[2pt]
Credit & 18,150 & 11 & \textbf{-1.28 \scriptsize{(0.01)}} & 10 & -0.92 \scriptsize{(0.01)} & \textbf{19} & -1.17 \scriptsize{(0.01)} & 4 & -1.17 \scriptsize{(0.01)} & 1 \\
\bottomrule
\end{tabular}

\end{table}

\subsection{Real and Synthetic Data Fit and Efficiency}\label{sec:bench}

We now assess the performance of TNBbeta regression models on real and synthetic data. Compared to beta regression, TNBbeta models have not only a different likelihood but also a different inference framework---a conditionally conjugate Gibbs sampler rather than HMC. Therefore, we contrast the predictive fit and computational efficiency of each approach.

\textbf{Data.} We fit each model to five real and two synthetic datasets. Both synthetic datasets are generated from a beta regression model, one with orthogonal and the other highly collinear predictors. See~\Cref{sec:app_exp} for additional experimental details and results.

\textbf{Models.} We fit two variants each of the TNBbeta and beta regression models from the previous section. In the first variant, the concentration is shared across observations---$q \sim \textrm{beta}(1,1)$ for TNBbeta and $\phi \sim \textrm{lognormal}(0,1)$ for beta. In the second, we regress covariates on the concentration---$q_i \teq \textrm{logit}^{-1}\big(X_i^\top\beta_q\big)$ for TNBbeta and $\phi_i \teq \exp\!\big(X_i^\top \beta\big)$ for beta. We also compare to cobin and micobin models in the constant concentration setting with default prior settings~\citep{lee2026scalable}. We specify each coefficient as $\beta_i \sim \mathcal{N}(0,1)$.

\textbf{Implementation details.} We fit each model $20$ times to each dataset, with $10$\% held-out, collecting $500$ samples after $500$ burn-in across $4$ MCMC chains on an NVIDIA A40 GPU.

\textbf{Results.}~\Cref{tab:bench_single} reports information rate and effective samples per second for each model with shared concentration on each dataset. The TNBbeta regression model provides a substantially better fit for three out of five real datasets. HMC-based beta regression produces the most effective samples per second on three of five real datasets, but its effective sample rate suffers when predictors are correlated, unlike TNBbeta models and the other data augmentation approaches. Furthermore,  TNBbeta models produce more effective samples per second than cobin and micobin models across all settings, suggesting that the data augmentation strategy underlying the TNBbeta is generally more efficient than that for the cobin models. We also note that all models attain similar predictive fit on the collinear synthetic dataset even though the data is truly beta-distributed.

The comparison between TNBbeta and beta regression when covariates are included for the concentration parameters is largely the same and the results are provided in~\Cref{tab:bench_dispreg}. Moreover, we note that the results in~\Cref{tab:bench_single} are with standardization of all covariates, which may not always be desired for interpreting regression coefficients. We find that posterior inference for beta regression via HMC can be significantly slowed without standardization, which can be seen in~\Cref{tab:bench_nostd}. In contrast, inference for the TNBbeta is largely unaffected.

\subsection{Spatial Models}\label{sec:trees}

In this section we show that the conditionally conjugate inference framework for TNBbeta models allows for easy extensions that accommodate more complex prior structure.

\begin{figure}[t]
\begin{center}
\includegraphics[width=\linewidth]{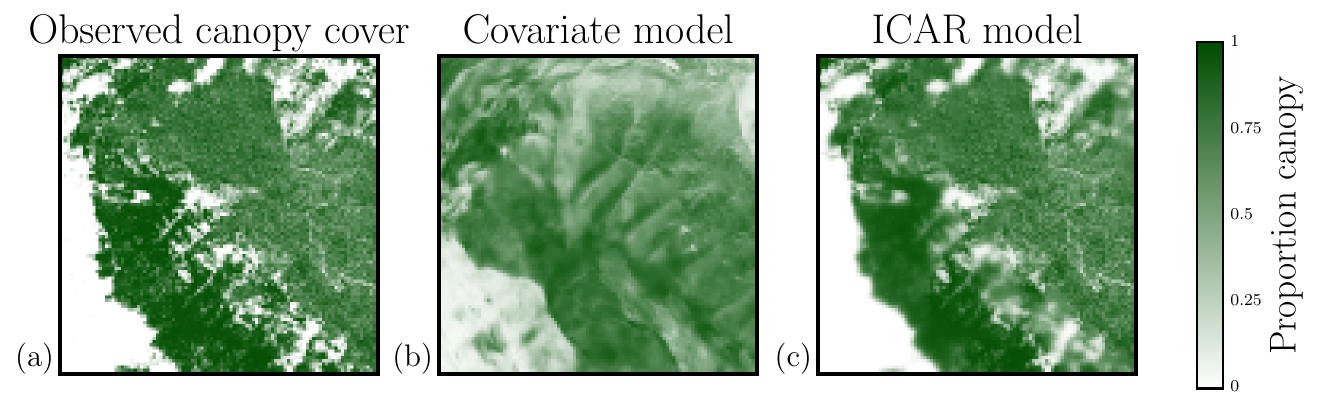}
\end{center}
\caption{TNBbeta models easily incorporate spatial structure. (a) Observed canopy cover. Predictive means for $p_i$ for the (b) covariate-only and (c) spatial TNBbeta model. \label{fig:canopy}}
\end{figure}

\textbf{Data.} We analyze a dataset of areal tree canopy derived from LiDAR scans of the Upper Gunnison Watershed~\citep{drew_bayesian_2025}. We partition the forest into $10{,}000$ $20 \!\times\! 20$ meter cells and compute $Y_i$ as the proportion of each cell covered by tree canopy (\Cref{fig:canopy}a). To enable a comparison with beta regression, we discard all sites with $0$\% canopy for a dataset with $8{,}763$ sites. Each site is associated with average values of six environmental covariates.

\textbf{Models.} We fit a suite of models to the canopy data. We specify each TNBbeta model as
\begin{equation}\label{eq:spatial_tnb}
\begin{aligned}
    Y_i &\sim \TNB\big(p_i,\, q_i,\, \varepsilon\big)\\
    p_i = \textrm{logit}^{-1}&\big(\psi_i^{(p)}\big), \qquad q_i = \textrm{logit}^{-1}\big(\psi_i^{(q)}\big),
\end{aligned}
\end{equation}
with different encoded structure $\psi_i^{(p)}$ and $\psi_i^{(q)}$ for the median and concentration, respectively, with fixed $\varepsilon \teq 1.0$. We fit two covariate-only models in which $\psi_i^{(p)} = \beta_0^{(p)} + X_i^\top \beta^{(p)}$ where $X_i \in \mathbb{R}^6$ are the environmental covariates for site $i$. Furthermore, we fit two spatial models in which $\psi_i^{(p)} = \varphi_i$ where $\varphi_i$ is a spatial random effect which we model with an intrinsic conditional autoregressive (ICAR) prior---$\boldsymbol{\varphi} \sim \textrm{ICAR}(\tau^2)$---under which each $\varphi_i \mid \boldsymbol{\varphi}_{\exclude i}$ is Gaussian with mean equal to the average of its neighbors and variance $\tau^2 / n_i$, where $n_i$ is the number of neighbors of cell $i$. We do not incorporate environmental covariates in the median of these models due to observed spatial confounding. For each parameterization of $\psi_i^{(p)}$, we fit two model variants, one in which the concentration is modeled as shared across all sites $\psi_i^{(q)} = \beta_0^{(q)}$ and one in which we regress the covariates on the concentration---$\psi_i^{(q)} = \beta_0^{(q)} + X_i^\top \beta^{(q)}$. We compare each model variant to the corresponding beta regression variant in which  $Y_i \sim \textrm{beta}\big(\phi_i \mu_i,\,\,\phi_i (1-\mu_i)\big)$ where $\mu_i = \textrm{logit}^{-1}\big(\psi_i^{(p)}\big)$ and $\phi_i = \textrm{exp}\big(\psi_i^{(q)}\big)$. We place standard normal priors on each regression coefficient and model the spatial variance as $\tau^2 \sim \Gamma(1,1)$. Again, we also fit the analogous micobin regression models.

\begin{figure}[t]
\begin{center}
\includegraphics[width=\linewidth]{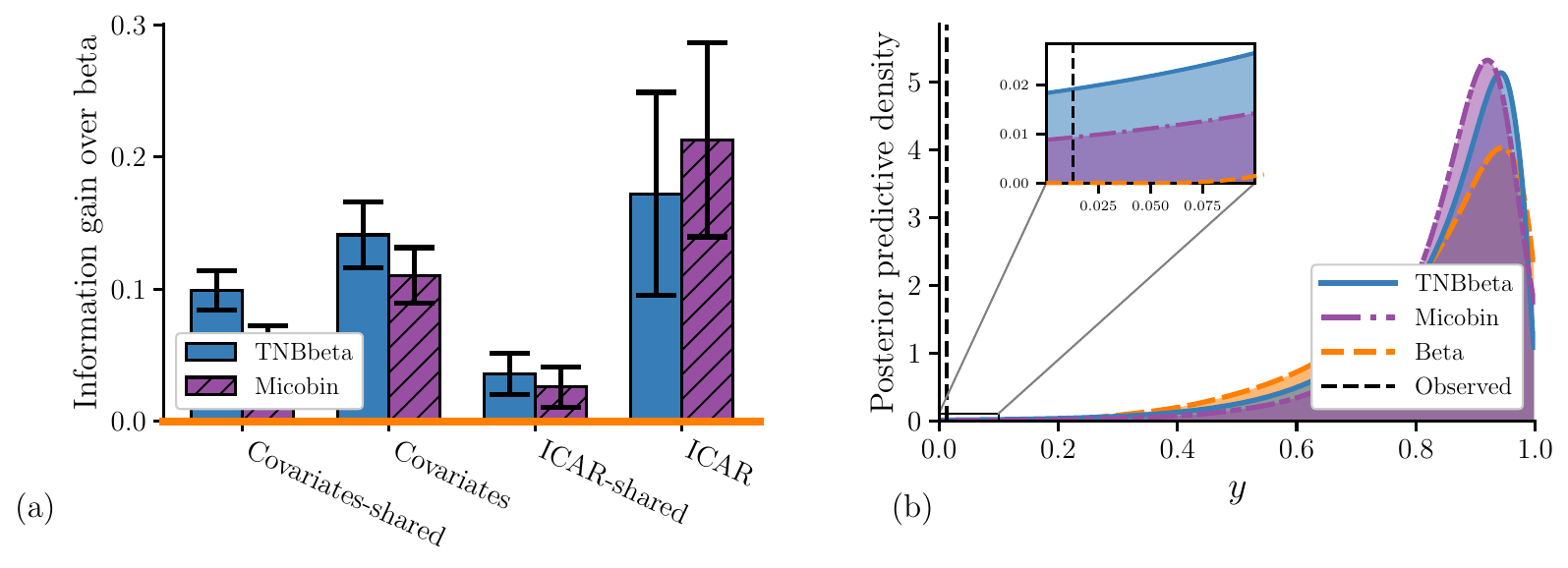}
\end{center}
\caption{(a) TNBbeta models predict canopy cover better than beta models across held-out sites, as shown by positive information gain $(\uparrow)$. (b) Example predictive distributions for a site in which the TNBbeta puts more density on an outlier observation.
\label{fig:gain}}
\end{figure}

\textbf{Implementation details.} We fit the covariate-only TNBbeta models with the Gibbs sampler from~\Cref{sec:gibbs}. For the spatial variants, we extend the sampler with ICAR random effects. Pólya-gamma augmentation renders the likelihood Gaussian in the linear predictor, pairing with the ICAR prior to yield a Gaussian conditional for $\boldsymbol{\varphi}$. We extend the Gibbs samplers provided by~\cite{lee2026scalable} to ICAR priors, which also utilize a Gaussian update after augmentation. We fit each beta model with Stan. We fit each model $20$ times, holding out $10 \!\times\! 10$ cell blocks, each time with $4$ chains, $2000$ burn-in, and $1000$ samples.

\textbf{Results.} \Cref{fig:gain}a shows information gain for each TNBbeta and micobin model over the equivalent beta model. The TNBbeta model typically outperforms the micobin model, but not always significantly. In each case, the TNBbeta and micobin model predict better than the beta model. One factor contributing to this superior predictive performance is the ability for the TNBbeta and micobin models to put nontrivial probability density near the boundary due to heavy-tailed behavior, which prevents catastrophic predictions for outlier observations. This phenomenon is illustrated for an example datapoint in \Cref{fig:gain}b.

\Cref{fig:canopy}b-c shows the posterior mean of each median $p_i$ for the covariate-only and spatial TNBbeta models fit to all $10,\!000$ canopy sites, including exact zeros. The spatial model under each likelihood captures far more structure than the environmental covariates can encode. Adding the spatial prior to the TNBbeta model does not increase computation, as both models fit in under $2$ minutes (see~\Cref{tab:tree_timing}). In contrast, fitting the spatial beta regression models via HMC incurs a severe computational cost, increasing from $8$ minutes to an hour on average for one variant. Moreover, while the micobin spatial model fits in $11$ minutes, computing the held-out likelihood is computationally intensive due to the lack of a closed-form density, requiring an additional $12$ minutes on average.

The availability of analytic forms and conditionally conjugate posterior inference makes the TNBbeta an appealing option for fitting Bayesian models of $[0,1]$-bounded support data, especially when outlier and zero observations are present. Moreover, Gaussian conditional conjugacy allows for practitioners to fit and critique increasingly complex models efficiently.

\section{Discussion}\label{sec:discuss}

We introduced the TNBbeta, a novel distribution on the unit interval parameterized by median $p$ and concentration $q$. This distribution is particularly well-suited for regression as both $p$ and $q$ can be modeled with covariates via logit link functions which admits simple, efficient, and tuning-free inference strategies via Pólya-gamma augmentation. This approach yields a flexible and robust class of models for bounded support data, including with exact boundary observations. The empirical studies in~\Cref{sec:robust,sec:eval_metrics,sec:bench} bear this out, demonstrating that TNBbeta regression models are often more sample-efficient, have better out-of-sample fit, and are more robust to outliers than either beta~\citep{ferrari_beta_2004} or micobin~\citep{lee2026scalable} regression models. Moreover, the case study in~\Cref{sec:trees} illustrates the ease with which complex or hierarchical TNBbeta models, in this case for spatial data, can be modularly constructed and efficiently fit.~\looseness=-1

The TNBbeta arises as a particular discrete mixture of the standard beta distribution expressible in terms of three negative binomial auxiliary variables. As discussed in~\Cref{sec:background}, this follows in a long line of generalized beta distributions defined by some discrete mixing distribution $\mathcal{DM}$, a tradition motivated by continuous mixtures of the beta being typically intractable~\citep{johnson_continuous_1995}. The generalized beta of~\citet{libby_multivariate_1982} for instance, arises from a negative binomial mixture over a single beta shape parameter, and is a special case of the TNBbeta, as we have shown. Another example is the DNCbeta~\citep{ongaro_results_2015,orsi2022new} defined by mixing over both shape parameters with Poisson random variables. Similar in spirit is the \textit{micobin} distribution, recently introduced by~\citet{lee2026scalable}, which is defined as a discrete mixture of the continuous binomial distribution. 

Discrete mixtures of bounded-support distributions are useful both for yielding more flexible model families and also for enabling tractable posterior inference schemes based on data augmentation. Such schemes, like the one we developed in~\Cref{sec:mcmc}, have been previously developed for inference in DNCbeta matrix factorization models~\citep{schein_doubly_2021,albert_doubly_2024} and for micobin regression~\citep{lee2026scalable}, among other examples.

Unlike previous approaches, TNBbeta models enjoy analytic properties that facilitate their use and interpretation. First, the TNBbeta density has a simple closed form, whereas the DNCbeta and micobin require costly infinite-series summation to evaluate. Second, inference for TNBbeta regression is tractable under the logit link, with coefficients interpretable as log-odds of the median or concentration. Finally, the conditional distribution of each auxiliary negative binomial under the TNBbeta is itself negative binomial, a remarkable instance of ``reverse conjugacy'', which makes posterior inference via data augmentation fast and easy to implement. In contrast, inference for DNCbeta and micobin models requires draws from non-standard families---i.e., the Bessel and Kolmogorov-gamma distributions, respectively---introducing a computational bottleneck and often necessitating approximations.

This work prompts several avenues of future research. The ``reverse conjugacy'' of the TNBbeta construction is itself a topic worthy of study. It relates to a family of beta generalizations introduced by~\cite{jones_slew_2023}, where the same conjugacy holds for a broad class of discrete priors over a single beta shape. Our results can be understood as an extension to the case of a dependent bivariate mixing distribution over both shape parameters. Characterizing more broadly when and how reverse conjugate constructions arise could enable inference in many new and interesting models. Another future avenue is in developing more hierarchical models based on the TNBbeta. A key benefit of the proposed approach is its modularity. Since latent variables retain conditionally Gaussian posteriors under data augmentation, the standard Gaussian toolkit invites temporal, hierarchical, sparsity-inducing, and other structured extensions. Finally, a natural topic of future study is in extending these ideas to the simplex, such as via discrete mixtures of the Dirichlet~\citep{orsi2025non}. Such extensions would enable new models for compositional data, as well as hierarchical constructions of simplex-constrained latent variables, both of which arise in many scientific applications.~\looseness=-1

{\spacingset{1.3}
\bibliographystyle{plainnat}
\bibliography{BayesianBetaRegression}}

\newpage
\appendix
\section*{Supplementary Materials}
\crefalias{section}{appendix}
\numberwithin{table}{section}
\numberwithin{figure}{section}
\input{sections/appendix}

\end{document}